\documentclass[pdflatex,sn-nature]{sn-jnl}

\usepackage{graphicx}%
\usepackage{multirow}%
\usepackage{amsmath,amssymb,amsfonts}%
\usepackage{amsthm}%
\usepackage{mathrsfs}%
\usepackage[title]{appendix}%
\usepackage{xcolor}%
\usepackage{textcomp}%
\usepackage{manyfoot}%
\usepackage{booktabs}%
\usepackage{algorithm}%
\usepackage{algorithmicx}%
\usepackage{algpseudocode}%
\usepackage{listings}%
\usepackage{tikz}
\usepackage{hyperref}
\usepackage{lineno}

\theoremstyle{thmstyleone}%
\theoremstyle{thmstyletwo}%

\theoremstyle{thmstylethree}%

\providecommand{\fnm}[1]{#1}
\providecommand{\sur}[1]{#1}
\providecommand{\orgdiv}[1]{#1}
\providecommand{\orgname}[1]{#1}
\providecommand{\city}[1]{#1}
\providecommand{\postcode}[1]{#1}
\providecommand{\country}[1]{#1}
\providecommand{\email}[1]{#1}

\providecommand{\orcid}[1]{}

\begin{document}
\title[Article Title]{Synchronization transitions and spike dynamics in a higher-order Kuramoto model with L\'{e}vy noise}


\author[1,2]{\fnm{Dan} \sur{Zhao}}\email{zzhaodannn@163.com}

\author[1,2]{\fnm{J\"urgen} \sur{Kurths}}\email{kurths@pik-potsdam.de}

\author[1]{\fnm{Norbert} \sur{Marwan}}\email{marwan@pik-potsdam.de}

\author*[3,4]{\fnm{Yong} \sur{Xu}}\email{hsux3@nwpu.edu.cn}

\affil[1]{\orgname{Potsdam Institute for Climate Impact Research}, \city{Potsdam}, \postcode{14412}, \country{Germany}}

\affil[2]{\orgdiv{Department of Physics}, \orgname{Humboldt University Berlin}, \city{Berlin}, \postcode{12489}, \country{Germany}}

\affil[3]{\orgdiv{School of Mathematics and Statistics}, \orgname{Northwestern Polytechnical University}, \city{Xi'an}, \postcode{710072}, \country{China}}

\affil[4]{\orgdiv{MOE Key Laboratory for Complexity Science in Aerospace}, \orgname{Northwestern Polytechnical University}, \city{Xi'an}, \postcode{710072}, \country{China}}

\abstract{Synchronization in various complex networks is significantly influenced by higher-order interactions combined with non-Gaussian stochastic perturbations, yet their mechanisms remain mainly unclear. In this paper, we systematically investigate the synchronization and spike dynamics in a higher-order Kuramoto model subjected to non-Gaussian L\'{e}vy noise. Using the mean order parameter, mean first-passage time, and basin stability, we identify clear boundaries distingusihing synchronization and incoherent states under different stability indexes and scale parameters of L\'{e}vy noise. For fixed coupling parameters, synchronization weakens as the stability index decreases, and even completely disappears when the scale parameter exceeds a critical threshold. By varying coupling parameters, we find significant dynamical phenomena including bifurcations and hysteresis. L\'{e}vy noise smooths the synchronization transitions and requires stronger coupling compared to Gaussian white noise. Furthermore, we define spikes and systematically investigate their statistical and spectral characteristics. The maximum number of spikes is observed at small scale parameter. A generalized spectral analysis further reveals burst-like structure via an edit distance algorithm. Moreover, a power-law distribution is observed in the large inter-window intervals of spikes, showing great memory effects. These findings deepen the understanding of synchronization and extreme events in complex networks driven by non-Gaussian L\'{e}vy noise.}

\maketitle


Synchronization is a universal phenomenon widely observed in natural and engineering systems, including neuronal brain networks~\cite{shahal2020synchronization}, biological rhythms~\cite{glass2001synchronization}, ecological networks \cite{blasius1999complex} and power grids~\cite{rohden2012self}. It provides insights into how collective order emerges from complex interactions among individual units~\cite{arenas2008synchronization,gallo2022synchronization}. The Kuramoto model serves as a fundamental mathematical framework for analyzing synchronization phenomena, describing the interactions among oscillators into a reduced and effective form. Its clear characteristics of synchronization transitions make it particularly beneficial for designing resilient networks and effectively suppressing oscillations in real world networks \cite{acebron2005kuramoto,rodrigues2016kuramoto, strogatz2000kuramoto,dorfler2012synchronization}.

However, the classical Kuramoto model considers only pairwise interactions, while many real-world systems rely on higher-order couplings, including triadic even higher-order interactions, significantly enriching the dynamical behaviors ~\cite{bick2023higher,benson2016higher,skardal2020higher}. Recent studies have identified several distinctive synchronization phenomena induced by higher-order couplings, such as abrupt desynchronization transitions, no synchronization, and multistability~\cite{skardal2019abrupt,ferraz2021phase}. Besides, Mill\'an et al.~\cite{millan2020explosive} present an explosive synchronization transition in a higher-order Kuramoto model. Clustering and abrupt desynchronization transitions are investigated arising from a three-way interaction in the Kuramoto model~\cite{xu2021spectrum}. Kundu et al.~\cite{kundu2022higher} demonstrates that higher-order interactions in nonlocally coupled identical Kuramoto oscillators promote the emergence of chimera states  without a nonzero phase lag. Complex spatiotemporal dynamics are illustrated on spatially embedded networks driven by local triadic interactions \cite{millan2024triadic}, resulting in abundant topological patterns. Furthermore, higher-order interactions have been shown to enhance the linear stability of synchronized states, while simultaneously shrinking their basins of attraction~\cite{zhang2024deeper}, revealing a fundamental trade-off between local stability and global robustness. These findings highlight the significant role of non-pairwise interactions in complex dynamical behaviors. Understanding these dynamics is important for practical applications, including the development of resilient power grids~\cite{albert2004structural}, the engineering of robust communication infrastructures~\cite{wang2021resilience,bian2025unveiling}, and the elucidation of intricate biological rhythms~\cite{goldbeter2008biological,dong2025adaptive}. 


In addition to interactions among oscillators, real systems are always influenced by stochastic excitation~\cite{zhao2023occurrence,wang2022coherence,zhang2021rate,wang2016levy}. Gaussian white noise is commonly used to model such external disturbances~\cite{campa2023synchronization,holder2017gaussian}. Many scholars have investigated the collective dynamical behaviors of higher-order oscillators under Gaussian white or colored noise, revealing richer dynamical phenomena compared to deterministic systems~\cite{zhao2023probabilistic, tyulkina2018dynamics,rajwani2025stochastic,marui2025synchronization}. We have previously investigated the heavy-tailed behavior of probability density functions (PDFs) in generalized Duffing systems driven by irregular periodic excitations under Gaussian colored noise, and found that stronger noise intensities and shorter correlation times lead to heavier tails~\cite{zhao2023probabilistic}. Recently, Marui et al.~\cite{marui2025synchronization} investigated synchronization bahaviors in higher-order Kuramoto oscillators with simplex interactions driven by Gaussian white noise. They found that even extremely weak noise erode synchronization states, leading to the lifetime of the synchronization state to exhibit an exponentially growing slow decay. However, realistic disturbances often exhibit characteristics different from Gaussian noise, such as large jumps, leading to rare but significantly impactful extreme events. L\'{e}vy noise, characterized by heavy-tailed distributions, naturally captures these extreme fluctuations and abrupt perturbations more accurately~\cite{zhao2025extreme,liu2023complex}. For example, Liu et al.~\cite{liu2023complex} studied stochastic transitions and stochastic resonance in a conceptual airfoil system subjected to non-Gaussian L\'{e}vy noise. 

Nonetheless, the synchronization dynamics of higher-order Kuramoto models under L\'{e}vy noise remains largely unexplored, representing a basic gap in our understanding of synchronization in complex networks with higher-order interactions. Moreover, L\'{e}vy noise often triggers extreme events, such as extreme synchronization events or spike events \cite{lee2025extreme,wang2022coherence}, which represent transient, high-coherence states significantly exceeding typical synchronization levels. These extreme events have profound implication for the stability and robustness of real-world systems. Analyzing these spikes provides critical insights into how heavy-tailed fluctuations can induce extreme behaviors in coupled oscillator systems.
Therefore, the present study investigates synchronization dynamics in higher-order Kuramoto models under L\'{e}vy noise, exploring how L\'{e}vy disturbances and higher-order interactions influence synchronization and spikes.

%
In this work, we show that Lévy fluctuations impede synchronization and stronger pairwise and higher-order interactions promote synchronization, revealing a critical interplay between higher-order interaction and non-Gaussian stochastic dynamics. Furthermore, we find that the amplitude and frequency of these spikes are not enhanced by stronger noise, but by lower Lévy noise scale parameter in conjunction with a larger stability index. These spikes also carry distinct statistical signatures of the underlying collective state, evident in their autocorrelation and power spectrum. Specifically, we find that the spike autocorrelation function decays rapidly in the incoherent regime, but exhibits significantly longer persistence during synchronization. Finally, the power spectra of these spiking events show power-law behavior, with underlying periodicities revealed by a windowed spectral analysis.
\section*{Results}
\section*{Model overview}
\label{sec:sec2}
We consider the following higher-order Kuramoto model subject to non-Gaussian L\'{e}vy noise. In particular, the phase dynamics of the $i$th oscillator is governed by \cite{suman2024finite}
\begin{linenomath}
\begin{equation}\label{eq:eq1}
	\dot{\theta}_{i}=\omega_{i}+\frac{K_{1}}{N}\sum_{j=1}^{N}\sin\left(\theta_{j}-\theta_{i}\right)+\frac{K_{2}}{N^{2}}\sum_{j=1}^{N}\sum_{k=1}^{N}\sin\left(2\theta_{j}-\theta_{k}-\theta_{i}\right)+\zeta\left(t\right),
\end{equation}
\end{linenomath}
in which $\theta_i$ is phase of the $i$th oscillator and $\omega_{i}$ is natural frequency of the $i$th oscillator. $N$ is the total number of oscillators. $K_1$ and $K_2$ are the coupling parameters of pairwise and triadic interaction, respectively. $\zeta(t)$ represents non-Gaussian L\'{e}vy noise. 

\section*{Basins of attraction in the noise-free case}\label{sec:sec7}
Considering fixed $K_1$ and $K_2$, we focus on the stable and unstable fixed point of the system (\ref{eq:eq1}). The nonlinear differential equation of the global order parameter $r(t)$ is as follows \cite{suman2024finite}.
\begin{linenomath}
	\begin{equation}\label{eq:eq4}
		\dot{r}=-r\Delta+\frac{K_1}{2}\left(r-r^3\right)+\frac{K_2}{2}\left(r^3-r^5\right).
	\end{equation}
\end{linenomath}

\begin{figure}[H]
	\centering
	\includegraphics[width=0.8\textwidth]{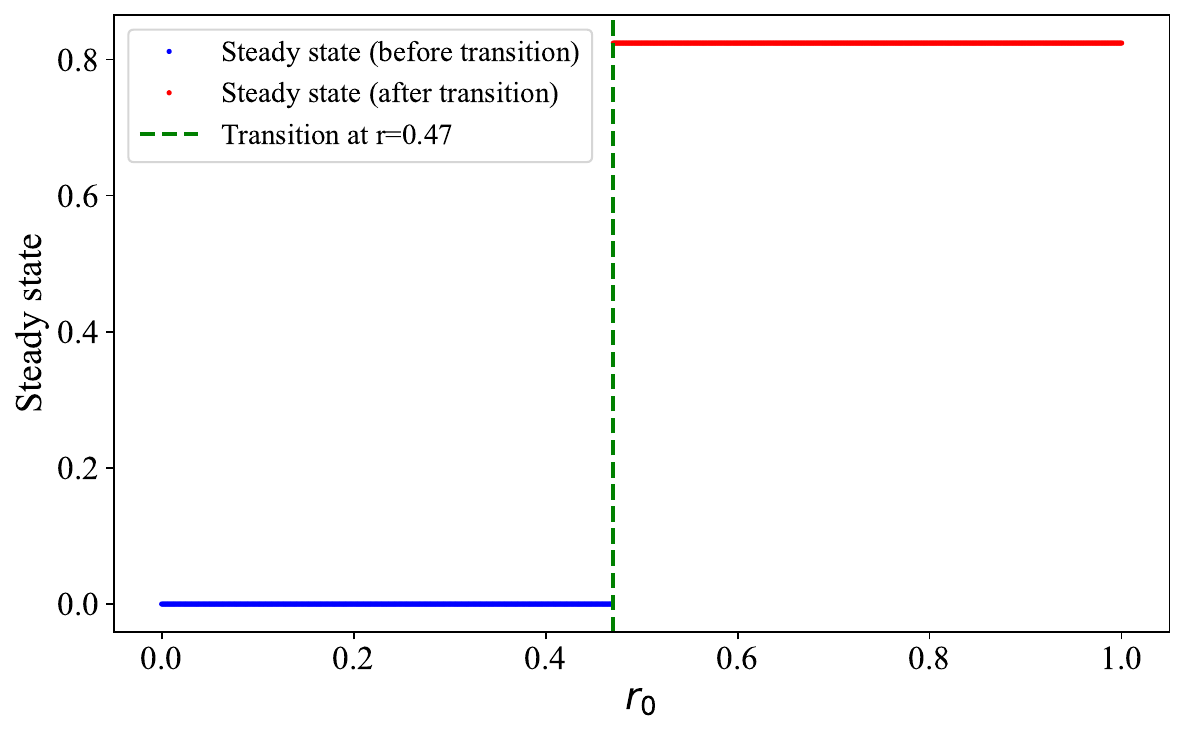}
	\caption{The stable and unstable points are attained by the Eq.~(\ref{eq:eq4}). Blue points describe $r=0$. Red points are $r=1$. $r_0=0.47$ represents the transition region.}
	\label{fig:fig1_1}      
\end{figure}

Equation (\ref{eq:eq4}) admits two stable states, namely $r=0$ and $r=1$. They represent incoherent state and fully synchronized state. By numerically integrating Eq.~(\ref{eq:eq4}) from various initial conditions $r_0\in[0,1]$, we obtain their basins of attraction, as shown in Fig.~\ref{fig:fig1_1}. Different initial conditions lead to two different states, including $r=0$ and $r=1$. Blue markers denote trajectories that converge to the incoherent state $r=0$, whereas red markers indicate those that approach full synchronization $r=1$. The vertical dashed line at $r_0=0.47$ marks the separatrix between these two basins. This sharp threshold underlines the system’s bistability. 

\begin{figure}
	\centering
	\includegraphics[width=0.9\textwidth]{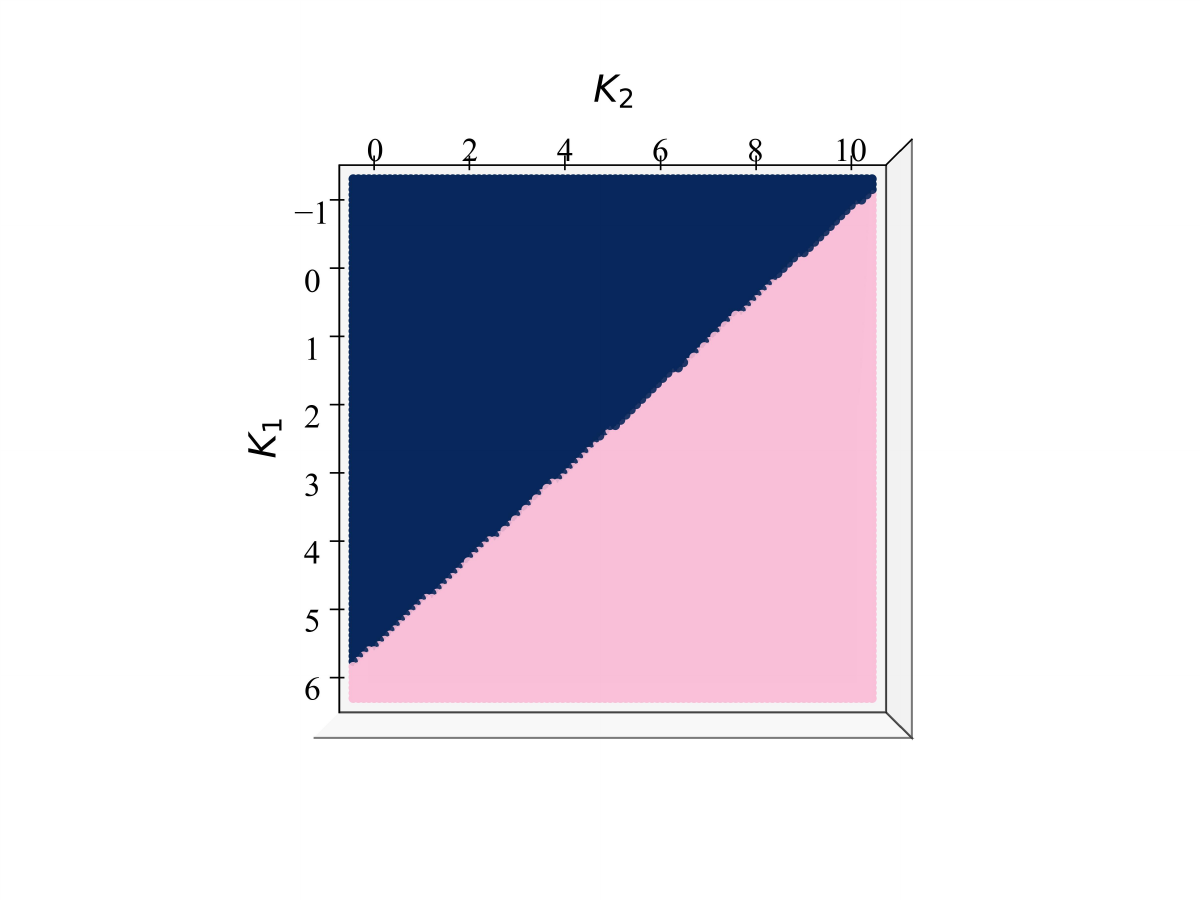}
	\caption{The basin of attraction under different $K_1$ and $K_2$. The initial conditions $r_{0}(t)\in[0,1]$. The pink and navy regions are the synchronization and incoherence, respectively. This is the top-down view of Supplementary Fig.~S1.}
	\label{fig:fig1_2}      
\end{figure}

In what follows, considering different $K_1$ and $K_2$, the basin of attraction will be calculated, as shown in Fig.~\ref{fig:fig1_2}. Bistable region occurs for different $K_1$ and $K_2$. The initial conditions are $r_0\in[0,1]$. If the long-time attractor is synchronized, $r\approx1$ as shown in pink. Besides, $r\approx0$ is the incoherent state, shown in navy. The perspective views of the three-dimensional basin of attraction are provided in Supplementary Fig.~S1, with the top-down view shown in Fig.~\ref{fig:fig1_2}.
We find from Fig.~\ref{fig:fig1_2} that all initial conditions $r_0$ flow to the synchronized state, when they lie above a certain tilted plane in the
$(K_1,K_2,r_0)$ space. This shows that sufficiently strong coupling guarantees global synchronization regardless of the initial coherence level of the population. Below this same plane, synchronization does not occur for any initial conditions as the incoherent attractor captures the entire cube defined by $r_0 \in [0,1]$. Especially, we find that $K_1+K_2\gtrsim 5$ guarantees synchronization, whereas weaker coupling, i.e., $K_1+K_2\lesssim 5$, can not overcome frequency disorder, leading to incoherence. Figure~S1 clearly supports this phenomenon from a different perspective.  

Based on the analysis of the steady states for different $K_1$ and $K_2$, we now proceed to investigate the synchronization transition of the system (\ref{eq:eq1}) driven by L\'{e}vy noise.

\section*{Synchronization transitions under fixed coupling parameters}
\label{sec:sec3}
\subsection*{Probability density functions of the order parameter}

\begin{figure}
	\centering
	\includegraphics[width=1\textwidth]{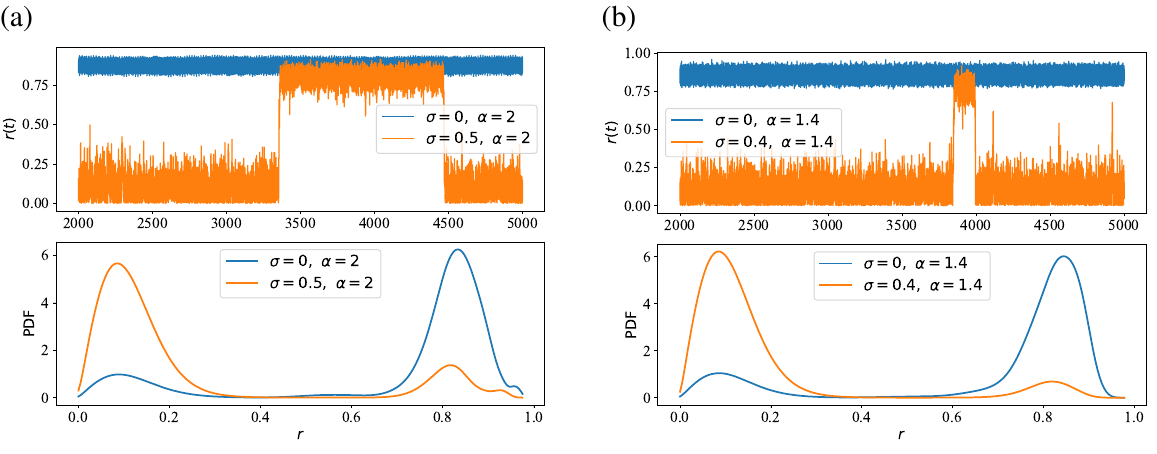}
	\caption{In each subfigure, the first row shows the time series of $r(t)$, and the second row shows the PDF obtained by multiple trajectories. Blue line respresents the deterministic case ($\sigma=0$). Orange line shows the stochastic case. (a) $\sigma=0.5$, $\alpha=2$ (orange lines) and $\sigma=0$, $\alpha=2$ (blue lines). (b) $\sigma=0.4$, $\alpha=1.4$ (orange lines) and $\sigma=0$, $\alpha=1.4$ (blue lines).}
	\label{fig:fig1_3}      
\end{figure}

We fix the coupling parameters at $K_1=0.8$, $K_2=8.0$. When noise effects are neglected, i.e., scale parameter $\sigma=0$, the system (\ref{eq:eq1}) may stabilize at either $r=0$ or $r=1$. This bistability, as illustrated in Fig.~\ref{fig:fig1_1}, results in the $r(t)$ shown by the blue curves in Fig.~\ref{fig:fig1_3}. Under the influence of noise, the responses on $r(t)$ corresponding to different values of stability index $\alpha$ are represented by the orange curves in Fig.~\ref{fig:fig1_3}. Besides, Fig.~\ref{fig:fig1_3} also illustrates the probability density functions (PDFs) of the global order parameter $r(t)$. The transient behaviors of the $r(t)$ have been removed. We infer from each figure that each PDF in the deterministic case is not exactly the same, which is caused by randomizing the initial conditions. 

From Fig.~\ref{fig:fig1_3}(a), when $\sigma=0$, $r(t)$ will eventually reach the synchronization state. Otherwise, $r(t)$ is more difficult to transition from $r=0$ to $r=1$, at $\sigma\ne0$, then returns to $r=0$ in a very short time, yielding a bimodal PDF with peaks near $r\approx0$ and $r\approx0.8$. In the intermediate case ($\alpha=1.4$), shown in Fig.~\ref{fig:fig1_3}(b), the system (\ref{eq:eq1}) exhibits rare and noise‐induced synchronization spikes, illustrating how heavy‐tailed perturbations can both suppress and transiently restore coherent behavior. The PDF, therefore, combines a dominant peak at $r\approx0$ with a small secondary peak near $r\approx0.8$. We observe a qualitatively similar transition for a wider range of $\alpha$ and $\sigma$, as detailed in the Supplementary Information (see Supplementary Fig.~S2).

Therefore, as 
$\alpha$ decreases, the same $\sigma$ both more readily suppresses persistent synchronization and occasionally gives rise to transient coherence events.  Whereas moderate Gaussian fluctuations can still drive the network from incoherence into synchrony, L\'{e}vy noise shifts the balance toward desynchronization. 

\begin{figure}[H]
	\centering
	\includegraphics[width=1.2\textwidth]{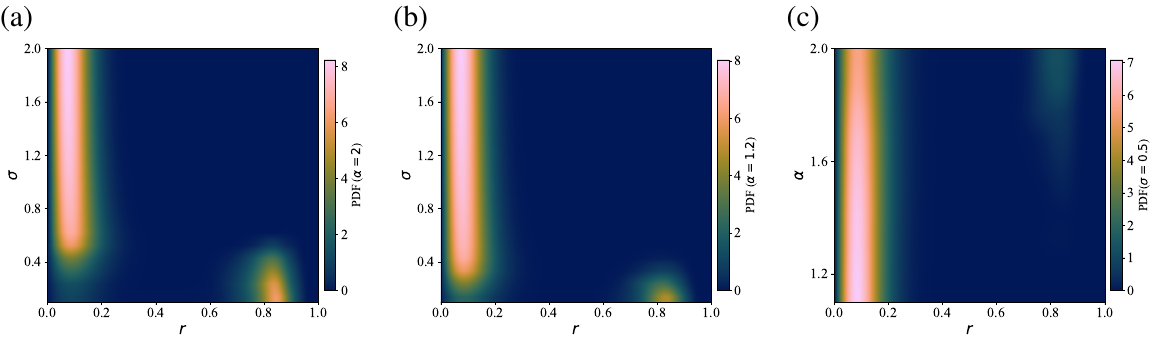}
	\caption{The PDF under different $\alpha$ and $\sigma$. The horizontal axis is the order parameter $r$, the vertical axis is the $\sigma$ and $\alpha$, and different colors represent PDF values. (a) $\alpha=2$. (b) $\alpha=1.2$. (c) $\sigma=0.5$.}
	\label{fig:fig1_4}      
\end{figure}

Furthermore, we consider the PDFs under different $\alpha$ and $\sigma$. Figures \ref{fig:fig1_4}(a)-(b) visualize how the stationary PDF of the order parameter $r$ is reshaped by $\sigma$. We plot the joint density $P(r,\sigma)$ for two fixed values of the stability index ($\alpha=2$ and $\alpha=1.2$), while panel (c) shows $P(r,\alpha)$ at a fixed $\sigma=0.5$. We find that, in each subfigure, two high‐density regions appear, one near $r\approx0.1$ (the incoherent attractor), and one near $r\approx0.8$ (the synchronized attractor).

As $\alpha$ decreases from 2.0 to 1.2, the incoherent peak both broadens in $\sigma$ and loses weight relative to the synchronized peak, namely heavy-tailed fluctuations progressively erode the stability of the synchronized branch. Besides, as the decrease of $\alpha$, $\sigma$ required to suppress synchronization decreases accordingly. Figures \ref{fig:fig1_4}(a)(b) verify them. We also provide the PDFs under $\sigma=0.5$ in Fig.~\ref{fig:fig1_4}(c). The system (\ref{eq:eq1}) mainly focus on the incoherent state, i.e., $r\approx0$. Therefore, the critical value of $\alpha$ needs to be determined based on the different values of $\sigma$. However, no matter what the value of $\alpha$ is, $\sigma>0.5$ will basically not result in synchronization. The PDF for the case of $\alpha=1.8$, presented in Supplementary Fig.~S3, further corroborates this phenomenon. In the subsections that follow, we will systematically quantify these phenomena in terms of the mean order parameter, mean first‐passage times, and basin stability, to map out the critical boundary $(\alpha_c,\sigma_c)$ beyond which synchronization is suppressed. When $\alpha < 1$, the first-order moment of L\'{e}vy noise does not exist, and, hence, we do not consider the case of $\alpha < 1$ here.

\subsection*{Basin stability and quantitative analysis of the synchronization transition}

\begin{figure}[H]
	\centering
	\includegraphics[width=1\textwidth]{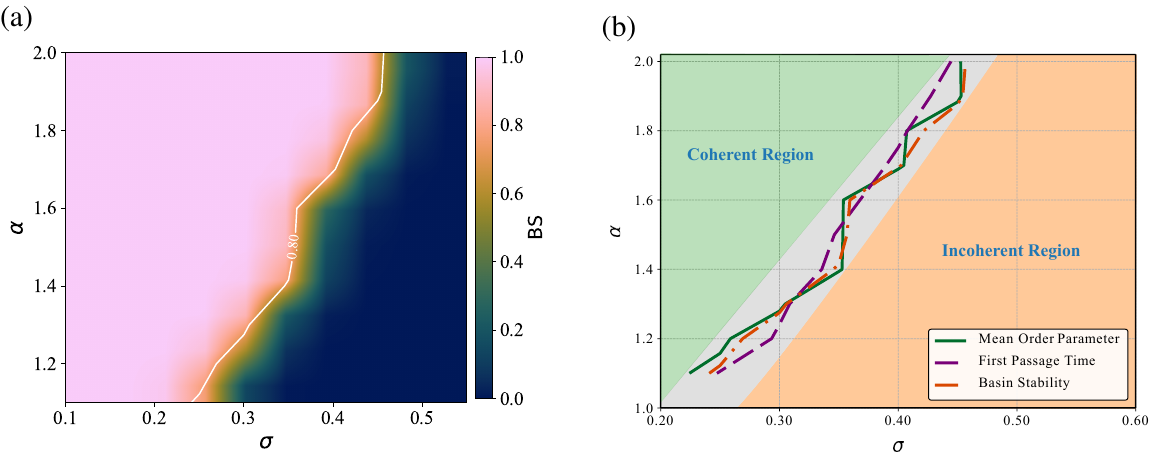}
	\caption{Heatmap of three metrics under different $\alpha$ and $\sigma$. (a) Basin stability $\mathrm{BS}$. Dark blue denotes $\mathrm{BS}\approx0$, almost no initial conditions synchronize. While pale pink denotes $\mathrm{BS}\approx1$, nearly all trajectories synchronize. The white curve is the contour line, marking the region with $\mathrm{BS}=0.8$. (b) Synthesis of three key metrics, including basin stability, mean order parameter, and MFPT.}
	\label{fig:fig1_7}      
\end{figure}

The results of basin stability $\mathrm{BS}$ (defined in the "Methods" section) are shown as Fig.~\ref{fig:fig1_7}(a) under different $\sigma$ and $\alpha$. We observe that, at fixed $\sigma$, increasing $\alpha$ raises $\mathrm{BS}$. In other words, when the noise has lighter tails, it is less effective at preventing synchronization. By contrast, for smaller $\alpha$, $\mathrm{BS}$ drops more rapidly, indicating that rare large jumps more easily trap the sytem (\ref{eq:eq1}) in the incorherent basin. Besides, at fixed $\alpha$, increasing $\sigma$ monotonically decreases $\mathrm{BS}$. Once $\sigma$ exceeds the white line, the majority of initial conditions fail to reach synchronization within the observation window. In practice, this means that Gaussian-like fluctuations require larger $\sigma$ to suppress synchronization than heavy-tailed L\'{e}vy fluctuations do.

Furthermore, the mean of the order parameter $\left< r \right>$ (defined by the "Methods" section) and the mean first-passage time (MFPT) $\mathrm{T}_{\text{first}}$ (illustrated by the "Methods" section) distribute at the similar transition region, including coherent and incoherent regions. The full results are provided in Supplementary Fig.~S4. To quantitatively characterize this transition, we simultaneously analyze three metrics, including $\mathrm{BS}$, $\left< r \right>$, and $\mathrm{T}_{\text{first}}$, with the results synthesized in Fig.~\ref{fig:fig1_7}(b). We find that the three contour critical curves have overlapping regions but are not identical. The discrepancy arises because the three indicators measure the dynamical behaviors from different perspectives, including the mean order parameter, the escape time, and the size-volume of basin. Even though the same threshold criteria are applied, the indicators focus on different aspects, leading to boundary deviations. Therefore, we define the largest region covered by the overlapping segments of the three curves as the transition region. Within this region, $\mathrm{BS}$ and $\left< r \right>$ may not simultaneously beyond 0.8 under the same parameters, and $T$ may not necessarily increase beyond $T_{\mathrm{max}}$. Consequently, synchronization occurs in the green region, but is suppressed in the yellow region, and synchronization in the transition region depends on specific conditions.

The above analysis illustrates the synchronization dynamics of system (\ref{eq:eq1}) under L\'{e}vy noise excitation at fixed coupling parameters $K_1$ and $K_2$. Specifically, as the stability index $\alpha$ decreases, synchronization becomes increasingly suppressed. Furthermore, when the scale parameter $\sigma$ exceeds 0.5, synchronization is suppressed. We validate these conclusions through heatmap analyses of the mean order parameter, the mean first-passage time, and the basin stability.

\section*{Modulation of synchronization by coupling parameters}
\label{sec:sec4}
%
%
\subsection*{Bifurcation hysteresis}

\begin{figure}
	\centering
	\begin{tikzpicture}
		\node[anchor=south west, inner sep=0] (IMG)
		{\includegraphics[width=1.1\textwidth]{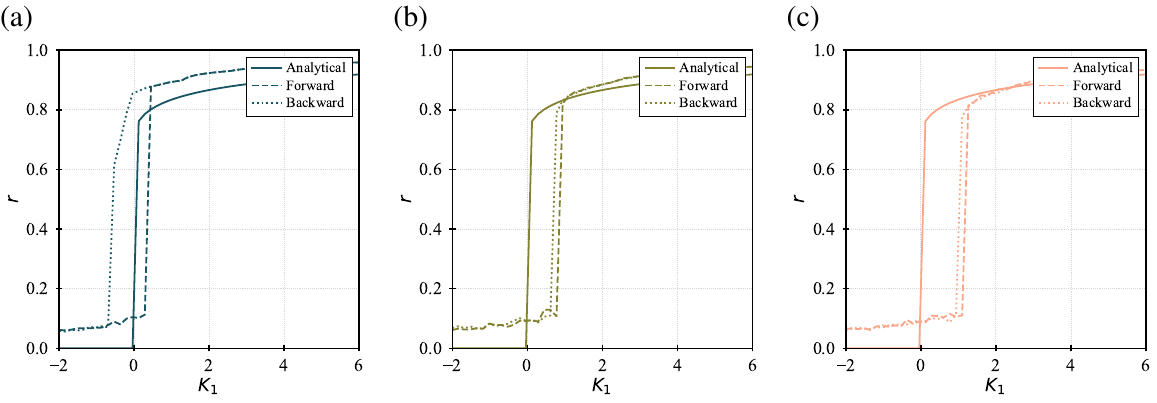}};
	\end{tikzpicture}
	\caption{The bifurcation diagrams under different $K_1$, $\alpha$ and $\sigma$. The solid line is the solution to Eq.~(\ref{eq:eq4}). Forward numerical iterations are denoted by long dashed lines with dots, while backward numerical iterations are denoted by short dashed lines. (a) The deterministic case, $\alpha=2$, $\sigma=0$. (b) The Gaussian case, $\alpha=2$, $\sigma=0.5$. (c) The L\'{e}vy noise, $\alpha=1.6$, $\sigma=0.5$.}
	\label{fig:fig1_9}
\end{figure}

In the following, varying $K_1$ is investigated. We set $K_2=8$ to describe the transition process. Figure \ref{fig:fig1_9} illustrates $r$ versus the first-order coupling $K_1$. The solid curve is the analytical solution of Eq.~(\ref{eq:eq4}), the long-dashed line with dots is the forward numerical simulation, i.e., increasing $K_1$. The short-dashed line is the backward simulation, i.e., decreasing $K_1$.

Figure \ref{fig:fig1_9}(a) refers to the deterministic case. A clear hysteresis loop opens, signifying bistability between the incoherent and synchronized states. Gaussian and L\'{e}vy noise case are compared in Figs.~\ref{fig:fig1_9}(b) and (c). In both stochastic cases, the overall bifurcation structure is preserved, but the curves shift to larger values of $K_1$, reflecting the extra coupling parameter needed to overcome noise and achieve synchrony. In addition, the case with Gaussian noise depicted in Fig.~\ref{fig:fig1_9}(b) shows a relatively small rightward shift, whereas the L\'{e}vy noise case in Fig.~\ref{fig:fig1_9}(c) exhibits a more pronounced shift to the right. In addition, we also generate bifurcation diagrams for different $K_2$, as shown in Supplementary Fig.~S5.

\subsection*{Integrated analysis of three metrics}
\begin{figure}
	\centering
	\includegraphics[width=1.2\textwidth]{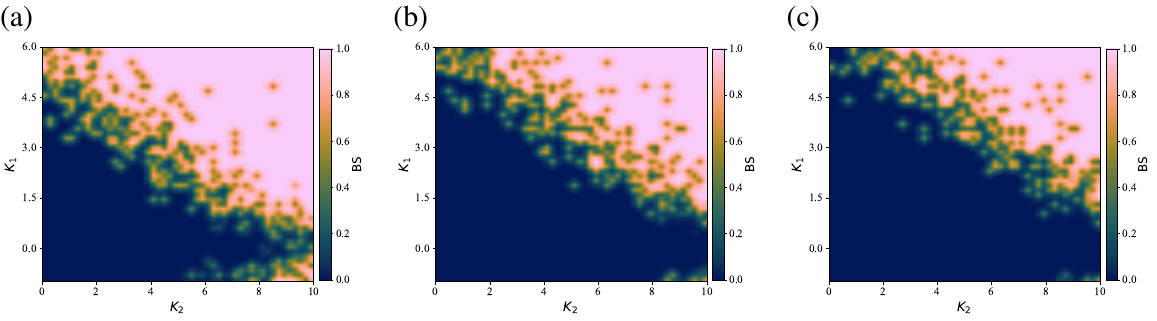}
	\caption{The basin stability $\mathrm{BS}$ under different $K_1$ and $K_2$. (a) The deterministic case. $\sigma=0$. (b) The Gaussian case. $\alpha=2$ and $\sigma=0.5$. (c) The L\'{e}vy case. $\alpha=1.6$ and $\sigma=0.5$.}
	\label{fig:fig1_12}      
\end{figure}

Furthermore, we now consider the $\mathrm{BS}$, the mean order parameter, and MFPT under varying $K_1$ and $K_2$. Figure \ref{fig:fig1_12} shows the $\mathrm{BS}$ as a function of $K_1$ and $K_2$ under different type of noise. In the deterministic case, the incoherent state dominates for low $K_1$ and low $K_2$,, i.e., $\mathrm{BS}=0$. Once $K_1$ and $K_2$ jointly exceed a boundary, the synchronization captures essentially the entire state, i.e., $\mathrm{BS}=1$. Considering Gaussian noise, in Fig.~\ref{fig:fig1_12}(b), adding moderate white noise fluctuations shifts the transition region to larger values of $K_1$ and $K_2$. In other words, stochastic perturbations make the synchronization less resilient. Besides, in Fig.~\ref{fig:fig1_12}(c), L\'{e}vy noise further reduces the resilience of the synchronization regime. 

\begin{figure}
	\centering
	\includegraphics[width=1.2\textwidth]{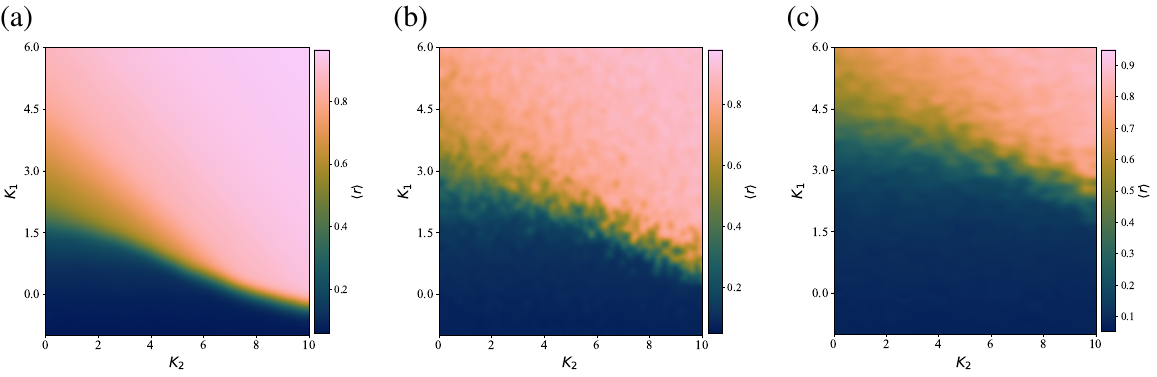}
	\caption{The mean order parameter $\left< r \right>$ under different $K_1$ and $K_2$. (a) The deterministic case. $\sigma=0$. (b) The Gaussian case. $\alpha=2$ and $\sigma=0.5$. (c) The L\'{e}vy case. $\alpha=1.6$ and $\sigma=0.5$.}
	\label{fig:fig1_10}      
\end{figure}

Figure \ref{fig:fig1_10} maps out the $\left< r \right> $ in the $(K_2,K_1)$ plane in three regimes, including the deterministic case, the Gaussian white noise and L\'{e}vy noise.
In Fig.~\ref{fig:fig1_10}(a), the system (\ref{eq:eq1}) exhibits a sharp transition in the absence of noise. Below a critical boundary in the $(K_2,K_1)$ space, the only attractor is the incoherent state. While the fully synchronized state dominates above the boundary. This produces the almost binary colormap. From Fig.~\ref{fig:fig1_10}(b), Gaussian fluctuations blur the threshold of sharp transition. Rather than an abrupt jump, $\left< r \right>$ rises gradually from 0 to 1 as the combined coupling of $K_1$ and $K_2$ increases. In the sub-threshold regime, weak but nonzero coherence level emerges, reflecting noise-induced excursions into the synchronous basin, i.e., $\left< r \right>\approx 0.2-0.6$. Beyond the blurred boundary, the system (\ref{eq:eq1}) still attains high coherence, but only $K_1$ and $K_2$ together can overcome the typical Gaussian fluctuation. Driven by non-Gaussian L\'{e}vy noise, in Fig.~\ref{fig:fig1_10}(c), the smoothing effect is even more pronounced. Meanwhile, the region of synchronization shrinks and the transition contour shifts to large $K_1$ and $K_2$. Rare but large jumps both facilitate escapes from incoherence and knock the system out of the synchronous basin, so that $\left< r \right>$ remains relatively low under moderate $K_1$ and $K_2$. Only when the couplings are strong enough to dominate these intermittent jumps does the system (\ref{eq:eq1}) spend most of its time near $r\approx 1$.


\begin{figure}
	\centering
	\includegraphics[width=1\textwidth]{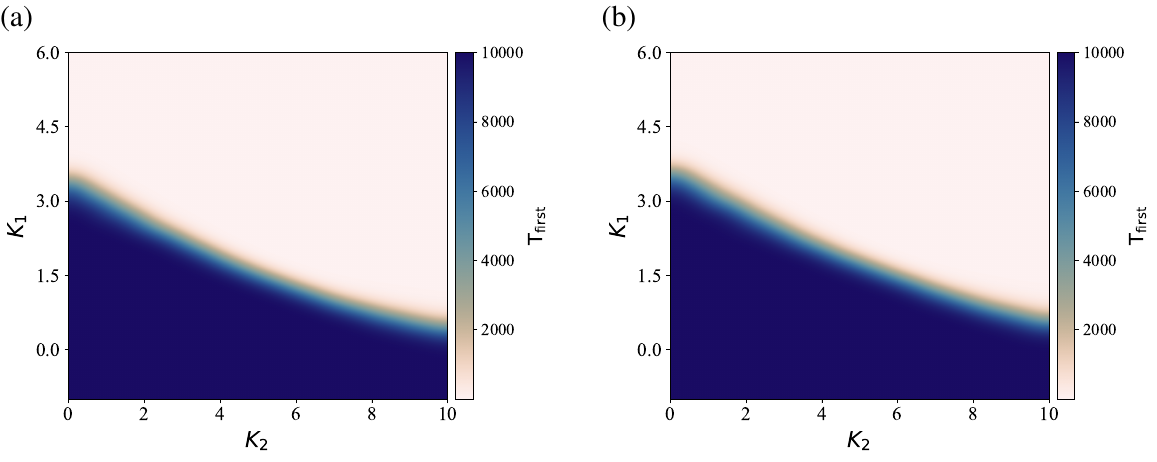}
	\caption{The MFPT $\mathrm{T}_{\text{first}}$ under different $K_1$ and $K_2$. (a) The Gaussian case. $\alpha=2$ and $\sigma=0.5$. (b) The L\'{e}vy case. $\alpha=1.6$ and $\sigma=0.5$.}
	\label{fig:fig1_11}      
\end{figure}
Figure \ref{fig:fig1_11} summarizes the MFPT $\mathrm{T}_{\text{first}}$. Figure \ref{fig:fig1_11}(a) is the result for Gaussian white noise, and Figure \ref{fig:fig1_11}(b) represents the L\'{e}vy noise. In both regimes, there is a clear boundary in the $(K_2,K_1)$ space. For weak coupling, i.e., low $K_1$ and low $K_2$, the $\mathrm{T}_{\text{first}}$ diverges. Beyond a roughly linear boundary, however, it sharply decreases. By comparsion, Figs.~\ref{fig:fig1_11}(a) and (b) show that L\'{e}vy noise systematically exhibits larger $\mathrm{T}_{\text{first}}$ than the Gaussian case for the same $(K_2,K_1)$.

Therefore, noise transforms the deterministic bistable transition into a broad crossover region. Gaussian fluctuations blur the boundary, but still allow an extensive region of synchronization. In contrast, L\'{e}vy fluctuations not only blur but also shift this boundary, requiring stronger coupling to reach synchronization. These results highlight that large jumps play an important role in shaping the collective dynamics of higher-order Kuramoto oscillators.

\section*{Spike statistics and spectral analysis}\label{sec:sec6}
As demonstrated in the above sections, the system (\ref{eq:eq1}) exhibits transitions between incoherence and synchronization. Particularly under the influence of L\'{e}vy noise, such transitions occur repeatedly and intermittently, driving the system (\ref{eq:eq1}) to alternate rather unpredictably between synchronization and incoherence. These intermittent and abrupt transitions can be regarded as extreme events in the dynamical system \cite{zhao2022extreme}. Therefore, it is important to analyze the frequency and characteristics of these recurrent extreme events, capturing the complex interplay between synchronization and incoherence.
\subsection*{Spike amplitude and the number of spikes}
\begin{figure}
	\centering
	\includegraphics[width=1.1\textwidth]{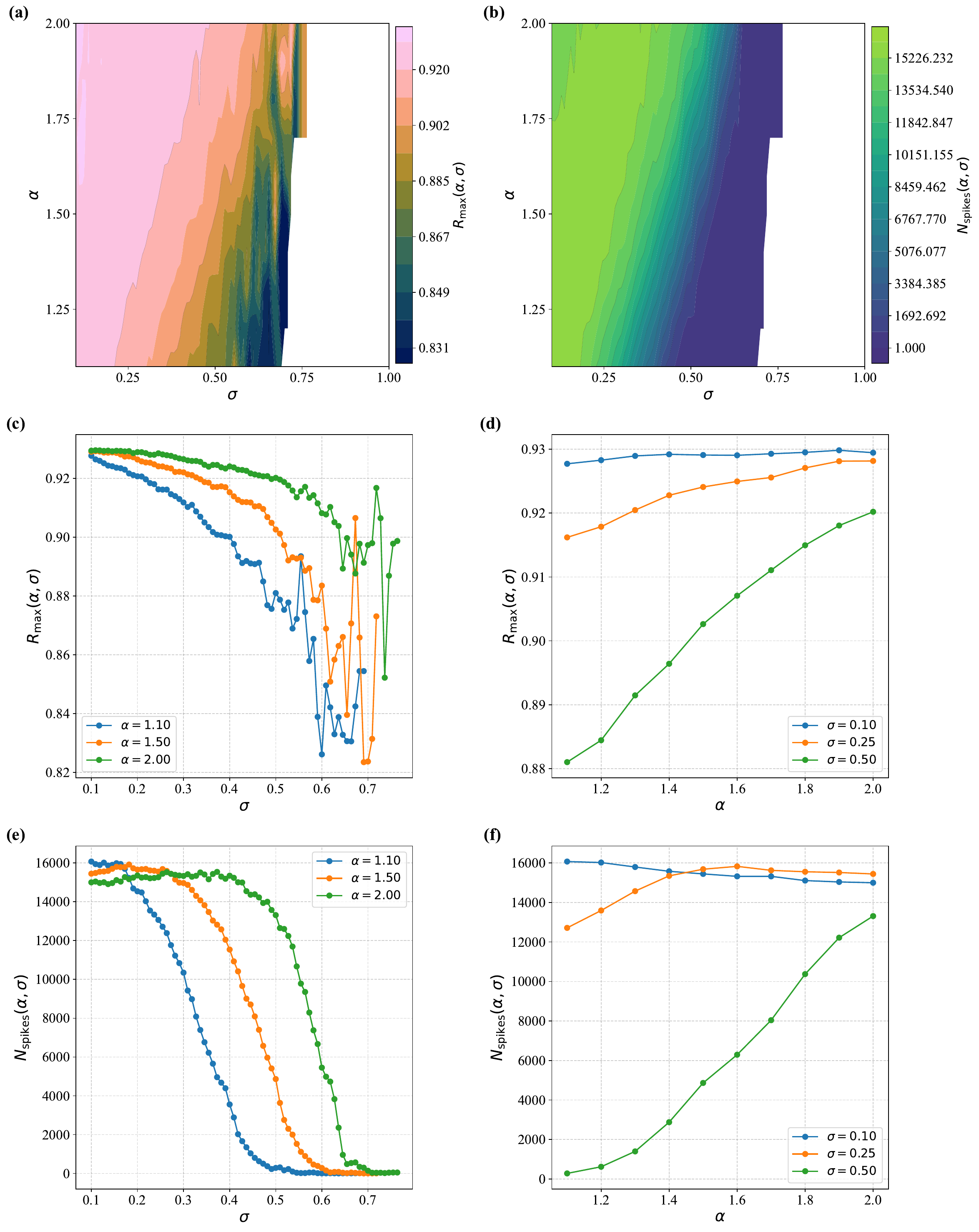}
	\caption{(a) The $R_{\mathrm{max}}(\alpha,\sigma)$ under different $\alpha$ and $\sigma$. Blank regions indicate no spikes occurred. (b) The $N_{\mathrm{spikes}}(\alpha,\sigma)$. Blank regions also indicate no spikes occurred. (c) The $R_{\mathrm{max}}(\alpha,\sigma)$ is plotted as a function of $\sigma$ for different $\alpha$. (d) $R_{\mathrm{max}}(\alpha,\sigma)$ is plotted as a function of $\alpha$ for different $\sigma$. (e) $N_{\mathrm{spikes}}(\alpha,\sigma)$ as a function of $\sigma$. Different colors correspond to different values of the parameter $\alpha$. (f) $N_{\mathrm{spikes}}(\alpha,\sigma)$ plotted against $\alpha$. Curves with different colors correspond to different $\sigma$.}
	\label{fig:fig1_14}      
\end{figure}

We quantify these spikes or extreme bursts using two metrics: amplitude and frequency. The results of maximum spike amplitude $R_{\mathrm{max}}(\alpha,\sigma)$ and spike count $N_{\mathrm{spikes}}(\alpha,\sigma)$ are shown in Fig.~\ref{fig:fig1_14}.


Figure \ref{fig:fig1_14}(a) depicts the averaged results of $R_{\mathrm{max}}(\alpha,\sigma)$ by multiple trajectories for different $\alpha$ and $\sigma$. It is worth noting that these averaged results are taken only when exceeding $r_{\mathrm{critical}}$, rather than each amplitude. We observe that when $\sigma$ is large, synchronization is suppressed completely, resulting in no spikes. In contrast, when $\sigma$ is small, $R_{\mathrm{max}}(\alpha,\sigma)$ becomes relatively large. These trends are displayed in Fig.~\ref{fig:fig1_14}(c). Although there are small fluctuations around $\sigma\in[0.6,0.8]$, $R_{\mathrm{max}}(\alpha,\sigma)$ exhibits an overall decreasing trend as $\sigma$ increases. Moreover, larger $\alpha$ are associated with a slower decline of $R_{\mathrm{max}}(\alpha,\sigma)$, whereas smaller $\alpha$ produce a faster decrease. Figure \ref{fig:fig1_14}(d) shows that $R_{\mathrm{max}}(\alpha,\sigma)$ increases with $\alpha$. This increase is stronger at larger $\sigma$ and negligible at small $\sigma$. These results illustrate the effect of L\'{e}vy noise on the maximum spike amplitude under different $\alpha$ and $\sigma$. In particular, smaller $\alpha$ as well as larger $\sigma$ promote stronger suppression of spikes. Figure \ref{fig:fig1_14}(b) shows the $N_{\mathrm{spikes}}(\alpha,\sigma)$ for different $\alpha$ and $\sigma$. Similar to Fig.~\ref{fig:fig1_14}(a), $N_{\mathrm{spikes}}(\alpha,\sigma)$ is highest when $\sigma$ is small and $\alpha$ is large, and vice versa. Moreover, when $\sigma$ is sufficiently large, the system (\ref{eq:eq1}) does not undergo transitions, resulting in no spikes. Besides, Figure \ref{fig:fig1_14}(e) shows a detailed $N_{\mathrm{spikes}}(\alpha,\sigma)$ under different $\alpha$. We observe that $N_{\mathrm{spikes}}(\alpha,\sigma)$ consistently decreases as $\sigma$ increases. Thus, smaller $\sigma$ corresponds to larger $N_{\mathrm{spikes}}(\alpha,\sigma)$. $N_{\mathrm{spikes}}(\alpha,\sigma)$ under different $\sigma$ are shown in Fig.~\ref{fig:fig1_14}(f). With increasing $\alpha$, $N_{\mathrm{spikes}}(\alpha,\sigma)$ either remains nearly constant or gradually increases. $N_{\mathrm{spikes}}(\alpha,\sigma)$ attains higher values for larger $\alpha$. Therefore, the system (\ref{eq:eq1}) always exhibits more frequent spikes when $\alpha$ is larger and $\sigma$ is smaller.


\subsection*{Autocorrelation and power spectrum analysis of spikes}
Moreover, analyzing only the amplitude and number of spikes is insufficient to fully characterize the impact of L\'{e}vy noise on synchronization transitions in the system. Therefore, we further explore the intrinsic temporal structural features of the spike sequences. Using the edit distance method~\cite{banerjee2020recurrence}, we compute the autocorrelation function in the time domain and the power spectrum in the frequency domain of the spike sequences. 

Spike sequences are discrete and irregularly spaced, even occasionally have large intervals. Therefore, it is difficult to directly calculate the autocorrelation function and power spectrum. We use the statistical properties of the autocorrelation function (EDACF) and power spectral estimate (EDSPEC) based on the edit distance \cite{marwan2023power}. The main idea behind the edit distance method is to quantify how similar two sequences of symbols or events are by calculating the smallest number of operations needed to turn one sequence into the other. These steps include adding, removing, and moving symbols \cite{marwan2023challenges}. 

\begin{figure}
	\centering
	\includegraphics[width=1\textwidth]{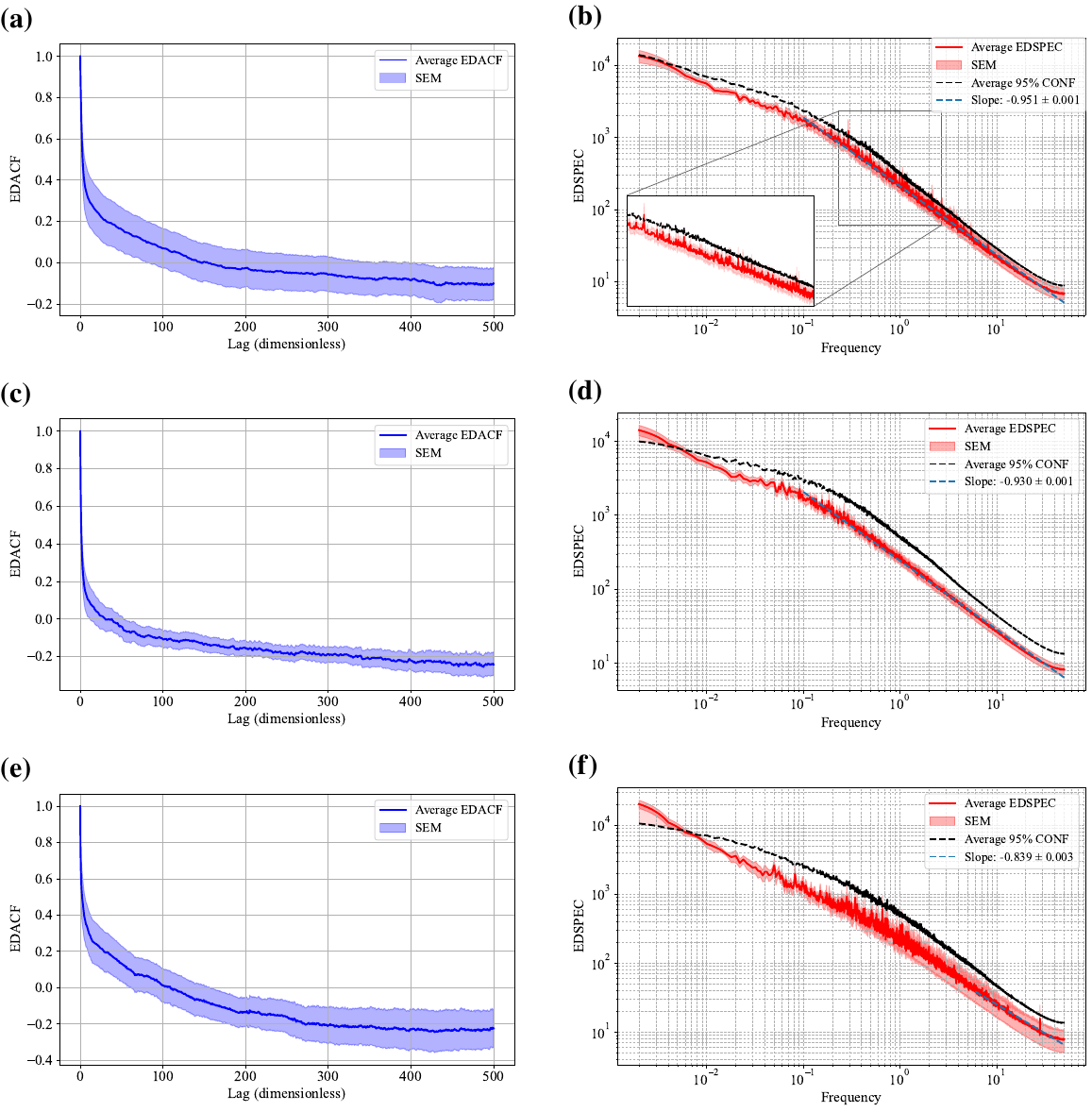}
	\caption{(a)-(b) $\alpha=1.6$, $\sigma=0.3$. (c)-(d) $\alpha=1.4$, $\sigma=0.33$. (e)-(f) $\alpha=1.4$, $\sigma=0.4$. (a)(c)(e) Estimated EDACF. (b)(d)(f) Estimated EDSPEC. EDSPEC on a log-log scale. The blue shaded region in EDACF indicates the SEM. In EDSPEC, the red shaded region also represents the SEM, while the black and green dashed lines show the 95$\%$ confidence interval and the least-squares fit to a suitable range, respectively.}
	\label{fig:fig1_15}      
\end{figure}

We select representative points from three distinct regions in Fig.~\ref{fig:fig1_7}(b) to investigate the autocorrelation and power spectra of the corresponding spike sequences. Specifically, the coherent region is characterized by $\alpha=1.6, \sigma=0.3$ (Figs.~\ref{fig:fig1_15}(a) and (b)), the transition region by $\alpha=1.4, \sigma=0.33$ (Figs.~\ref{fig:fig1_15}(c) and (d)), and the incoherent region by $\alpha=1.4, \sigma=0.4$ (Figs.~\ref{fig:fig1_15}(e) and (f)). The blue and red shaded regions represent one standard error of the mean (SEM), while the black curves indicate the $95\%$ confidence levels (see the "Methods" section).

We observe from Fig.~\ref{fig:fig1_15}(a) that the EDACF decays to zero at approximately lag values greater than 100, whereas in Fig.~\ref{fig:fig1_15}(c), the EDACF decays more rapidly, reaching zero at lag values smaller than 100. In Fig.~\ref{fig:fig1_15}(e), the EDACF reaches zero at around lag$\approx$100. These findings indicate that spike sequences in the coherent region display the longest-lasting serial dependency, whereas sequences in the incoherent region exhibit the shortest temporal correlations. Furthermore, from Figs.~\ref{fig:fig1_15}(b), (d), and (f), the spike sequences consistently follow a power-law $f^{-0.951}$, $f^{-0.930}$, and $f^{-0.839}$, respectively. Then we give the hypothesis: $H_0$: The temporal structure of the observed event sequence is indistinguishable from that of a random process, implying the absence of periodicity or long-range correlation. $H_1$: The event sequence exhibits a significant non-random temporal structure, such as periodicity, long-range correlation, or notable autocorrelation (see the "Methods" section). If the EDSPEC of the actual data exceeds these thresholds at certain frequencies at the $95\%$ confidence level, we reject the null hypothesis, indicating a significant non-random temporal structure in the system (\ref{eq:eq1}). Specifically, the EDSPEC in Figs.~\ref{fig:fig1_15}(d) and (f) remains within the $95\%$ confidence intervals, indicating no significant periodicities. However, there are peaks in the EDSPEC in Fig.~\ref{fig:fig1_15}(b), which exceed the $95\%$ confidence interval, indicating the existence of real but small spectral peaks.

\begin{figure}[H]
	\centering
	\includegraphics[width=1.1\textwidth]{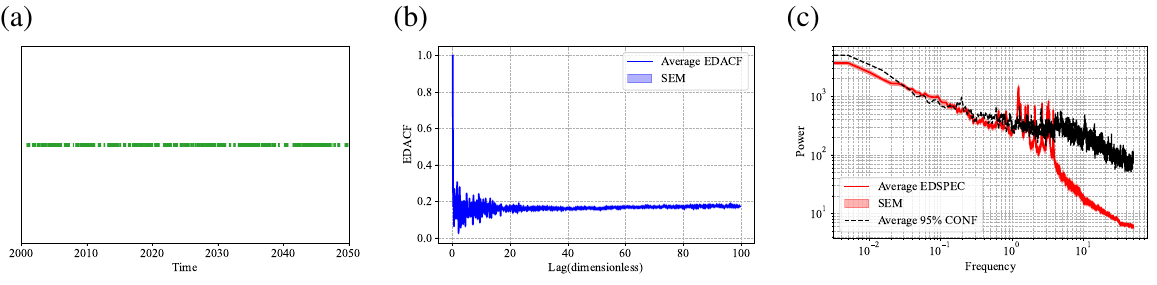}
	\caption{Analysis of spike sequences segmented into multiple windows. $\alpha=1.1$ and $\sigma=0.1$. (a) The segment of the spike sequence after window segmentation. (b) Averaged EDACF computed across multiple segmented windows, with shading indicating the SEM. (c) Averaged EDSPEC calculated across multiple segmented windows, also displaying SEM as a shaded area. Black curve shows the $95\%$ confidence interval.}
	\label{fig:fig1_16_1}      
\end{figure}

Considering spike sequences characterized by large inter-event intervals, it becomes more meaningful to analyze these sequences by segmenting them into shorter windows. Thus, we present the EDACF and EDSPEC calculated for windowed spike segments, allowing us to examine the temporal variations and local characteristics within the spike sequences in more detail. We assume that spikes are grouped into sequences if intervals between spikes exceeded a threshold $I_{\mathrm{threshold}}=200$. Each segment is, thus, bounded by intervals greater than $I_{\mathrm{threshold}}$. Then the EDACF and EDSPEC are calculated independently within each window. Taking the case of $\alpha=1.1$ and $\sigma=0.1$ as an example, the averaged EDACF and EDSPEC results from multiple segmented windows are shown in Fig.~\ref{fig:fig1_16_1}.

Figure \ref{fig:fig1_16_1}(a) shows an example segment of spikes over a time window, illustrating the discrete and irregular spikes. Figure \ref{fig:fig1_16_1}(b) shows the averaged EDACF computed by multiple segmented windows. The EDACF decays rapidly from an initial lag and stabilizes at a relatively low value, indicating limited long-range autocorrelation across spike events. Shaded regions represent the SEM. It means that the averaged EDACF lies entirely within the shaded SEM regions, indicating a high consistency across segmented windows. Thus, the observed spike dynamics are robust. Figure \ref{fig:fig1_16_1}(c) displays the averaged EDSPEC, which reveals a power-law behavior at lower frequencies, followed by distinct peaks at higher frequencies. Besides, multiple clear peaks exceed the 95\% confidence interval, which was established based on a null model of random spiking activity. This result allows us to reject the null hypothesis, confirming that the observed oscillations are statistically significant. The low variability, demonstrated by the SEM completely encompassing the EDSPEC curve, further indicates the presence of robust, burst-like structures. Of course, we test on multiple trajectories to obtain EDACF and EDSPEC, showing similar phenomena.

\begin{figure}[H]
	\centering
	\includegraphics[width=1.1\textwidth]{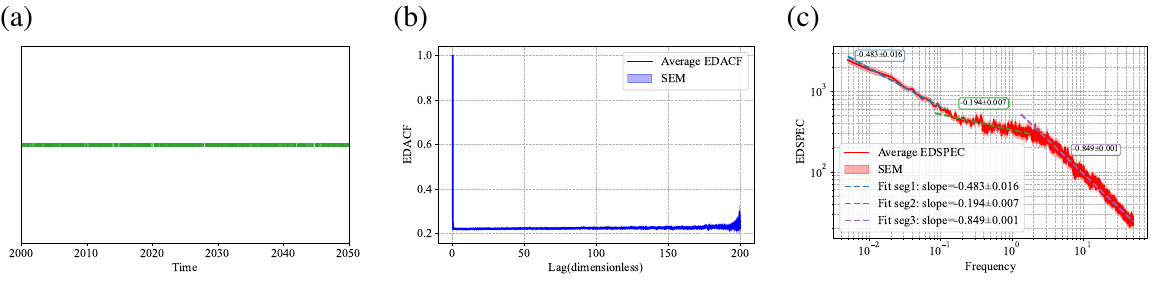}
	\caption{Analysis of L\'{e}vy noise spike sequences using window segmentation. $\alpha=1.1$ and $\sigma=0.1$. (a) Segment of the spike sequences extracted from L\'{e}vy noise. (b) Averaged EDACF computed over segmented windows, with SEM indicated by shading. (c) Averaged EDSPEC calculated across segmented windows, also with SEM depicted by shading. The three dashed lines are the fitted curves for the three frequency bands, respectively.}
	\label{fig:fig1_16_2}      
\end{figure}

For comparison, we also present the L\'{e}vy sequence, along with the estimated EDACF and EDSPEC. Notably, the previously threshold $r_{\mathrm{critical}}$ is unsuitable for this L\'{e}vy sequence. We adopt the 80$\%$ percentile as an appropriate threshold to obtain spikes. Similarly, we divide the sequence into windows of length 200 to obtain the series shown in Fig.~\ref{fig:fig1_16_2}(a), the spike sequence appears relatively uniform. Hence, we directly apply the edit distance to estimate the average EDACF and EDSPEC across multiple windows. The results are illustrated in Figs.~\ref{fig:fig1_16_2}(b) and (c), respectively. EDACF shows a sharp initial peak rapidly decreasing to a consistently low level. It indicates that the spikes generated by L\'{e}vy noise exhibit rather weak correlations. When the lag is large, the fluctuation is small. In Fig.~\ref{fig:fig1_16_2}(c), EDSPEC shows a steady declining power spectrum, highlighting that L\'{e}vy noise exhibits low-frequency behavior with a simple monotonic decay. In addition, we fit three frequency bands in Fig.~\ref{fig:fig1_16_2}(c). All three bands show a power law decay, with $f^{-\beta_1}$, $f^{-\beta_2}$, and $f^{-\beta_3}$. $\beta_1=0.483$, $\beta_2=0.194$, and $\beta_3=-0.849$, which indicates long-range correlation and stationarity.

\begin{figure}
	\centering
	\includegraphics[width=1\textwidth]{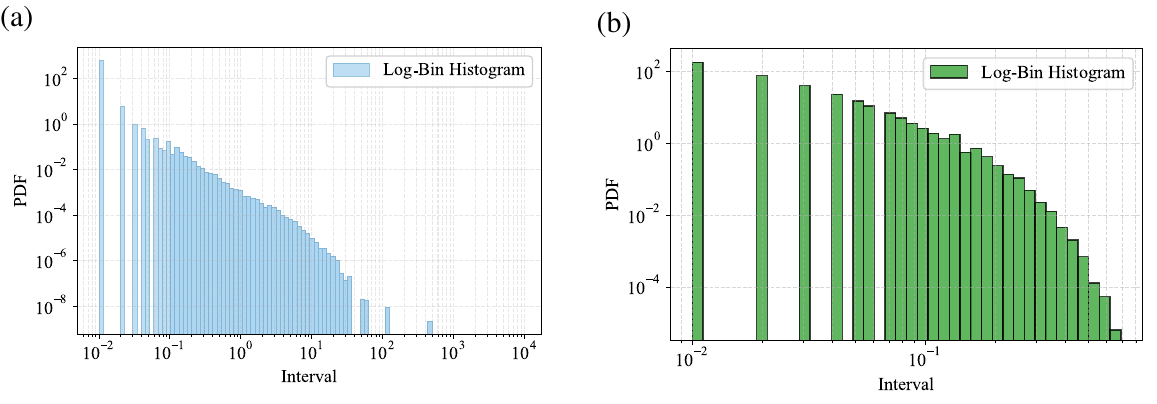}
	\caption{PDFs of spike intervals for system (\ref{eq:eq1}). (a) The histogram of spike intervals combined with short intra-window intervals and longer inter-window intervals. A power-law fit characterizes the heavy-tailed behavior. (b) The PDF of spike intervals, without requiring window segmentation due to shorter inter-spike intervals.}
	\label{fig:fig1_16_3}      
\end{figure}

To demonstrate the fundamental difference in temporal correlation between the collective dynamics of the system (\ref{eq:eq1}) and the L\'{e}vy noise, we compare the spike interval distributions of the order parameter $r(t)$ and $\zeta(t)$. The distribution for the spike sequences of $r(t)$ exhibits a clear power-law decay over several orders of magnitude (Fig.~\ref{fig:fig1_16_3}(a)). This behavior is a hallmark of long-range correlations, indicating that events in the system's output are not independent but possess a long memory. In contrast, the interval distribution for $\zeta(t)$ decays much faster than a power law, showing a progressively steeper slope on a log-log scale (Fig.~\ref{fig:fig1_16_3}(b)). This demonstrates a lack of long-range correlation. Therefore, this comparison confirms that the complex, long-memory dynamics observed in the system (\ref{eq:eq1}) are an emergent property, rather than a simple reflection of the input noise characteristics.

\section*{Discussion}
This study reveals that L\'{e}vy noise actively suppresses synchronization in higher-order Kuramoto models, standing in direct opposition to the classic phenomenon of noise-induced transitions and stochastic resonance. Our findings suggest that the interaction between the heavy-tailed L\'{e}vy noise and the higher-order interactions is key, which effectively destroy new synchronized groups before they grow. In the desynchronized state, the spikes are not random and isolated events. Instead, their temporal clustering into bursts suggests a residual, transient form of local organization. The quasi-periodic nature suggested by the burst-like structure could correspond to a characteristic timescale of this formation-and-destruction process of transient synchronous clusters.

This suppression mechanism is not merely a modeling curiosity but suggests a new principle, offering a potential mechanism for preventing runaway synchronization and epileptic seizures, as well as a dynamical explanation for polarization in socioeconomic systems. Hypersynchronization is often linked to pathological states like epilepsy~\cite{lepeu2024critical,li2025synchronization}. The brain also employs powerful mechanisms to maintain a healthy, desynchronized state essential for complex computation. Our model could be interpreted as a representation of such a regulatory mechanism. The higher-order connections represent complex neural assemblies, while L\'{e}vy-like excitations could model strong, disruptive external stimuli or internal channel noise. Thus, the noise-induced suppression of synchronization could be a protective feature, preventing runaway synchronization and seizures. The observed bursts might then correspond to transient, localized neural computations that are performed without triggering pathological global synchronization. 

Besides, consensus formation in social groups or stability in financial markets often relies on complex, many-body interactions~\cite{baumann2020modeling}. Our findings suggest that in the presence of randomly recurring major shocks, such as political scandals and market crashes, it becomes extremely difficult to form and maintain a global consensus or stable state. The system thus remains trapped in a desynchronized regime. Within this state, only transient, small-scale bursts could form before being shattered by the next crisis, offering a dynamical explanation for the persistent polarization or volatility seen in real systems under stress.

Of course, there are several limitations in this study. Firstly, we conduct on a network with a simplified topology. The interplay of complex topologies, such as random network or small-world network, with higher-order structures and L\'{e}vy noise remains to be explored. Secondly, the parameters of the L\'{e}vy noise are explored in a fixed range. Exploring a wider range of parameters could reveal more complex dynamical phenomena.

Therefore, future work could proceed in several directions. A systematic investigation across the full parameter space of L\'{e}vy noise is needed to map out synchronization behaviors. Implementing the model on empirical higher-order networks, such as brain connectomes or collaboration networks, would be a critical step toward validating our theoretical findings. Finally, a deeper analysis of the burst dynamics, using tools from time-series analysis to formally quantify their periodicity and statistical properties, would help to confirm our hypothesis regarding transient cluster formation and provide a richer understanding of the complex behaviors.

\section*{Conclusion}

In this paper, we investigate the synchronization and spike dynamics of a higher-order Kuramoto model driven by L\'{e}vy noise. First, we analyze fixed coupling parameters $K_1$ and $K_2$, which correspond to pairwise and higher-order interactions, respectively, and identify how L\'{e}vy noise parameters significantly affect synchronization. Specifically, synchronization becomes suppressed with decreasing the L\'{e}vy stability index $\alpha$. Moreover, synchronization is always inhibited for a sufficiently large noise scale parameter $\sigma$, as confirmed by the basin stability, mean order parameter, and mean first-passage time. Next, by modulating $K_1$ and $K_2$, we explore bifurcation hystersis phenomena. The presence of L\'{e}vy noise transforms sharp deterministic synchronization boundaries into broader and smoother transitions, requiring stronger coupling to achieve comparable synchronization compared to Gaussian noise. Finally, we focus on spike dynamics induced by L\'{e}vy noise, defining spikes as periods during which the order parameter exceeds a predefined threshold. We calculate the maximize amplitude and the number of spikes. Both always occur at positions with a smaller scale parameter $\sigma$. Spectrum analysis is applied by an edit distance algorithm. By comparing spike intervals and L\'{e}vy noise intervals, we further emphasize power-law scaling introduced by higher-order Kuramoto under non-Gaussian noise. 

Crucially, these results illustrate that both synchronization transitions and spike occurrences under L\'{e}vy noise can be viewed as manifestations of extreme events in oscillator networks. Synchronization transitions represent collective extreme coherence phenomena, while spike dynamics correspond to transient, highly coherent states that substantially deviate from typical dynamics. Such unified understanding of extreme events induced by non-Gaussian noise expands our insights into synchronization dynamics, highlighting the critical role of higher-order interactions and L\'{e}vy-type fluctuations in governing rare but impactful dynamical behaviors. Our findings provide valuable perspectives for managing and predicting extreme events in real-world complex oscillator networks subjected to stochastic excitations.

\section*{Methods}
\subsection*{L\'{e}vy noise and model parameters}
L\'{e}vy noise is the formal derivative of the L\'{e}vy process $L(t)$ to time $t$. $L(t)$ is expressed by its characteristic function $\phi_{\zeta}(t)$~\cite{zan2022response}, namely
\begin{linenomath}
	\begin{equation}\label{eq:eq2}
		\left.\phi_{\zeta}(t)=\left\{
		\begin{array}
			{ll}\exp\left\{i\mu t-\sigma^{\alpha}|t|^{\alpha}\left[1-i\beta\operatorname{sgn}(t)\tan\left(\frac{\pi\alpha}{2}\right)\right]\right\}, & \alpha\in(0,1)\cup(1,2], \\
			\exp\left\{i\mu t-\sigma|t|\left[1+i\beta\operatorname{sgn}(t)\frac{2}{\pi}\ln|t|\right]\right\}, & \alpha=1.
		\end{array}\right.\right.
	\end{equation}
\end{linenomath}
In Eq.~(\ref{eq:eq2}), $\alpha$ is the stability index and $\beta$ the skewness index. When $\alpha=2$, L\'{e}vy noise degenerates into Gaussian noise. $\beta>0$ means a left-skewed PDF and $\beta<0$ is a right-skewed PDF. $\beta=0$ shows the symmetric case. $\mu\geqslant0$ is the location parameter and $\sigma$ the scale parameter. So the noise intensity is $D=\sigma^\alpha$.

For the system (\ref{eq:eq1}), we take $N=100$. Each natural frequency $\omega_{i}$ is sampled independently via a Cauchy distribution. Its expectation is $\omega_0=0$ and its standard deviation is $\Delta=1$. 
\subsection*{Order parameter and three metrics}
We calculate the global order parameter $r(t)$ by Eq.~(\ref{eq:eq3}).
\begin{linenomath}
	\begin{equation}\label{eq:eq3}
		z(t)=r(t)e^{\iota\Psi(t)}=\frac{1}{N}\sum_{j=1}^{N}e^{\iota\theta_i(t)},
	\end{equation}
\end{linenomath}
where $r(t)\in[0,1]$ measures the phase coherence and $\Psi(t)$ denotes the average phase. It can be calculated by the order parameters of multiple paths as $\left< r \right> =\left< \left| N^{-1}\sum_j{e^{\iota \theta _i}} \right| \right>$ and provides quantitative insights in the collective dynamics.

Basin stability is the probability that the system converges to a steady state under a specific initial condition, revealing a measure of the resilience against perturbations~\cite{menck2013basin}. We obtain some initial condition samples from phase space, calculate the probability that the order parameter goes into a synchronization state experiencing long enough time. 
\begin{linenomath}
	\begin{equation}
		\mathrm{BS}=\dfrac{n_{\mathrm{path}}}{N_{\mathrm{path}}},
	\end{equation}
\end{linenomath}
in which $n_{\mathrm{path}}$ represents the number of paths going into the synchronization state, described by $r>r_{\mathrm{threshold}}$. $N_{\mathrm{path}}$ is the number of all paths and $r_{\mathrm{threshold}}=0.8$. If $\mathrm{BS}$ is large, most of the initial values in system (\ref{eq:eq1}) will go into the synchronization state. Otherwise, the system (\ref{eq:eq1}) is very sensitive to initial values and could not go into the synchronization state. In Fig.~\ref{fig:fig1_7}(b), a white contour at $\mathrm{BS}=0.8$ highlights the boundary between mostly synchronized and mostly incoherent regimes, delineating the region where at least 80$\%$ of random initial conditions end up in the synchronized state.

For the mean of order parameter $\left< r \right>$, in Fig.~\ref{fig:fig1_7}(b), the contour $\left< r \right>=0.8$ marks the combinations of $(\alpha,\sigma)$ at which the long-time average $\left< r \right>$ falls below 0.8 or above 0.8. It characterizes the average synchronization or incoherence behaviors of the system (\ref{eq:eq1}) under noise fluctuation. 

First-passage time is defined as the ensemble-averaged time at which the order parameter $r$ first exceeds a chosen threshold $r_{\mathrm{threshold}}$~\cite{ma2022early,zan2022reliability}. Specifically,
\begin{linenomath}
	\begin{equation}
		\mathrm{T}_{\text{first}}=\left\langle\inf\left\{t:r\left(t\right)>r_{\mathrm{threshold}}\right\}\right\rangle,
	\end{equation}
\end{linenomath}
where $r_{\mathrm{threshold}}$ is the threshold we defined. According to the value of $r$ under different states, we choose $r_{\mathrm{threshold}}=0.8$. When $r\ge0.8$, the system (\ref{eq:eq1}) will enter the synchronization state. Besides, in Fig.~\ref{fig:fig1_7}(b), we identify a threshold $T_{\mathrm{threshold}}=0.8T_{\mathrm{max}}$ for MFPT, where $T_{\mathrm{max}}$ is the largest value of $T$ we considered.

\subsection*{Spike amplitude and the number of spikes}
Given the critical threshold $r_{\mathrm{critical}}=0.5$, we define a spike as a time interval during which $r(t)\geq r_{\mathrm{critical}}$. These intervals represent sudden synchronization events. Illustrations of the spikes are presented in the Supplementary Fig.~S6. To provide a comprehensive framework for assessing and suppressing extreme events, we calculate the maximum amplitude of spikes and the number of spikes during the specific time interval extracted for each $\alpha$ and $\sigma$. The maximum spike amplitude is defined as $R_{\mathrm{max}}(\alpha,\sigma)=\max_{t\in \left[ 0,T \right]}\{r(t)\mid r(t)\geqslant r_{\mathrm{critical}}\}$. The spike count is defined by 
\begin{linenomath}
	\begin{equation}\nonumber
		N_{\mathrm{spikes}}(\alpha,\sigma) = \#\left\{[t_{\mathrm{start}}, t_{\mathrm{end}}]\subseteq[0,T]\;:\; 
		r(t_{\mathrm{start}}^-)\!<\!r_{\mathrm{critical}},\; r(t_{\mathrm{end}}^+)\!<\!r_{\mathrm{critical}},\; r(t)\geq r_{\mathrm{critical}}\right\}.
	\end{equation}
\end{linenomath}
\subsection*{Edit distance algorithm}
When calculating the averaged EDACF and EDSPEC, we select multiple trajectories to generate spike sequences for each parameter set. The step size of lag is 0.01. To establish the $95\%$ confidence intervals shown, we generate 20 surrogate event sequences by randomly and uniformly redistributing events within the original observation interval. These surrogate datasets are then used to calculate the EDACF and EDSPEC, from which we derived frequency-specific thresholds corresponding to the $95\%$ significance level. These thresholds allow a statistical comparison with the original data to test the hypotheses.

\section*{Data availability}
All data needed to evaluate the findings of the paper are available within the paper.
\section*{Code availability}
The Python and Julia source code is available at \url{https://github.com/zhaodan-npu/Higher_order_kuramoto_paper_code}.

\bibliography{mybibfiles} 

\begin{thebibliography}{10}
\expandafter\ifx\csname url\endcsname\relax
  \def\url#1{\burl{#1}}\fi
\expandafter\ifx\csname urlprefix\endcsname\relax\def\urlprefix{URL }\fi
\providecommand{\bibinfo}[2]{#2}
\providecommand{\eprint}[2][]{\url{#2}}
\providecommand{\doi}[1]{\url{https://doi.org/#1}}
\bibcommenthead

\bibitem{shahal2020synchronization}
\bibinfo{author}{Shahal, S.} \emph{et~al.}
\newblock \bibinfo{title}{Synchronization of complex human networks}.
\newblock \emph{\bibinfo{journal}{Nature Communications}}
  \textbf{\bibinfo{volume}{11}}, \bibinfo{pages}{3854} (\bibinfo{year}{2020}).

\bibitem{glass2001synchronization}
\bibinfo{author}{Glass, L.}
\newblock \bibinfo{title}{Synchronization and rhythmic processes in
  physiology}.
\newblock \emph{\bibinfo{journal}{Nature}} \textbf{\bibinfo{volume}{410}},
  \bibinfo{pages}{277--284} (\bibinfo{year}{2001}).

\bibitem{blasius1999complex}
\bibinfo{author}{Blasius, B.}, \bibinfo{author}{Huppert, A.} \&
  \bibinfo{author}{Stone, L.}
\newblock \bibinfo{title}{Complex dynamics and phase synchronization in
  spatially extended ecological systems}.
\newblock \emph{\bibinfo{journal}{Nature}} \textbf{\bibinfo{volume}{399}},
  \bibinfo{pages}{354--359} (\bibinfo{year}{1999}).

\bibitem{rohden2012self}
\bibinfo{author}{Rohden, M.}, \bibinfo{author}{Sorge, A.},
  \bibinfo{author}{Timme, M.} \& \bibinfo{author}{Witthaut, D.}
\newblock \bibinfo{title}{Self-organized synchronization in decentralized power
  grids}.
\newblock \emph{\bibinfo{journal}{Physical Review Letters}}
  \textbf{\bibinfo{volume}{109}}, \bibinfo{pages}{064101}
  (\bibinfo{year}{2012}).

\bibitem{arenas2008synchronization}
\bibinfo{author}{Arenas, A.}, \bibinfo{author}{D{\'\i}az-Guilera, A.},
  \bibinfo{author}{Kurths, J.}, \bibinfo{author}{Moreno, Y.} \&
  \bibinfo{author}{Zhou, C.}
\newblock \bibinfo{title}{Synchronization in complex networks}.
\newblock \emph{\bibinfo{journal}{Physics Reports}}
  \textbf{\bibinfo{volume}{469}}, \bibinfo{pages}{93--153}
  (\bibinfo{year}{2008}).

\bibitem{gallo2022synchronization}
\bibinfo{author}{Gallo, L.} \emph{et~al.}
\newblock \bibinfo{title}{Synchronization induced by directed higher-order
  interactions}.
\newblock \emph{\bibinfo{journal}{Communications Physics}}
  \textbf{\bibinfo{volume}{5}}, \bibinfo{pages}{263} (\bibinfo{year}{2022}).

\bibitem{acebron2005kuramoto}
\bibinfo{author}{Acebr{\'o}n, J.~A.}, \bibinfo{author}{Bonilla, L.~L.},
  \bibinfo{author}{P{\'e}rez~Vicente, C.~J.}, \bibinfo{author}{Ritort, F.} \&
  \bibinfo{author}{Spigler, R.}
\newblock \bibinfo{title}{The kuramoto model: A simple paradigm for
  synchronization phenomena}.
\newblock \emph{\bibinfo{journal}{Reviews of Modern Physics}}
  \textbf{\bibinfo{volume}{77}}, \bibinfo{pages}{137--185}
  (\bibinfo{year}{2005}).

\bibitem{rodrigues2016kuramoto}
\bibinfo{author}{Rodrigues, F.~A.}, \bibinfo{author}{Peron, T. K.~D.},
  \bibinfo{author}{Ji, P.} \& \bibinfo{author}{Kurths, J.}
\newblock \bibinfo{title}{The kuramoto model in complex networks}.
\newblock \emph{\bibinfo{journal}{Physics Reports}}
  \textbf{\bibinfo{volume}{610}}, \bibinfo{pages}{1--98}
  (\bibinfo{year}{2016}).

\bibitem{strogatz2000kuramoto}
\bibinfo{author}{Strogatz, S.~H.}
\newblock \bibinfo{title}{From kuramoto to crawford: exploring the onset of
  synchronization in populations of coupled oscillators}.
\newblock \emph{\bibinfo{journal}{Physica D}} \textbf{\bibinfo{volume}{143}},
  \bibinfo{pages}{1--20} (\bibinfo{year}{2000}).

\bibitem{dorfler2012synchronization}
\bibinfo{author}{Dorfler, F.} \& \bibinfo{author}{Bullo, F.}
\newblock \bibinfo{title}{Synchronization and transient stability in power
  networks and nonuniform kuramoto oscillators}.
\newblock \emph{\bibinfo{journal}{SIAM Journal on Control and Optimization}}
  \textbf{\bibinfo{volume}{50}}, \bibinfo{pages}{1616--1642}
  (\bibinfo{year}{2012}).

\bibitem{bick2023higher}
\bibinfo{author}{Bick, C.}, \bibinfo{author}{Gross, E.},
  \bibinfo{author}{Harrington, H.~A.} \& \bibinfo{author}{Schaub, M.~T.}
\newblock \bibinfo{title}{What are higher-order networks?}
\newblock \emph{\bibinfo{journal}{SIAM Review}} \textbf{\bibinfo{volume}{65}},
  \bibinfo{pages}{686--731} (\bibinfo{year}{2023}).

\bibitem{benson2016higher}
\bibinfo{author}{Benson, A.~R.}, \bibinfo{author}{Gleich, D.~F.} \&
  \bibinfo{author}{Leskovec, J.}
\newblock \bibinfo{title}{Higher-order organization of complex networks}.
\newblock \emph{\bibinfo{journal}{Science}} \textbf{\bibinfo{volume}{353}},
  \bibinfo{pages}{163--166} (\bibinfo{year}{2016}).

\bibitem{skardal2020higher}
\bibinfo{author}{Skardal, P.~S.} \& \bibinfo{author}{Arenas, A.}
\newblock \bibinfo{title}{Higher order interactions in complex networks of
  phase oscillators promote abrupt synchronization switching}.
\newblock \emph{\bibinfo{journal}{Communications Physics}}
  \textbf{\bibinfo{volume}{3}}, \bibinfo{pages}{218} (\bibinfo{year}{2020}).

\bibitem{skardal2019abrupt}
\bibinfo{author}{Skardal, P.~S.} \& \bibinfo{author}{Arenas, A.}
\newblock \bibinfo{title}{Abrupt desynchronization and extensive multistability
  in globally coupled oscillator simplexes}.
\newblock \emph{\bibinfo{journal}{Physical Review Letters}}
  \textbf{\bibinfo{volume}{122}}, \bibinfo{pages}{248301}
  (\bibinfo{year}{2019}).

\bibitem{ferraz2021phase}
\bibinfo{author}{Ferraz~de Arruda, G.}, \bibinfo{author}{Tizzani, M.} \&
  \bibinfo{author}{Moreno, Y.}
\newblock \bibinfo{title}{Phase transitions and stability of dynamical
  processes on hypergraphs}.
\newblock \emph{\bibinfo{journal}{Communications Physics}}
  \textbf{\bibinfo{volume}{4}}, \bibinfo{pages}{24} (\bibinfo{year}{2021}).

\bibitem{millan2020explosive}
\bibinfo{author}{Mill{\'a}n, A.~P.}, \bibinfo{author}{Torres, J.~J.} \&
  \bibinfo{author}{Bianconi, G.}
\newblock \bibinfo{title}{Explosive higher-order kuramoto dynamics on
  simplicial complexes}.
\newblock \emph{\bibinfo{journal}{Physical Review Letters}}
  \textbf{\bibinfo{volume}{124}}, \bibinfo{pages}{218301}
  (\bibinfo{year}{2020}).

\bibitem{xu2021spectrum}
\bibinfo{author}{Xu, C.} \& \bibinfo{author}{Skardal, P.~S.}
\newblock \bibinfo{title}{Spectrum of extensive multiclusters in the kuramoto
  model with higher-order interactions}.
\newblock \emph{\bibinfo{journal}{Physical Review Research}}
  \textbf{\bibinfo{volume}{3}}, \bibinfo{pages}{013013} (\bibinfo{year}{2021}).

\bibitem{kundu2022higher}
\bibinfo{author}{Kundu, S.} \& \bibinfo{author}{Ghosh, D.}
\newblock \bibinfo{title}{Higher-order interactions promote chimera states}.
\newblock \emph{\bibinfo{journal}{Physical Review E}}
  \textbf{\bibinfo{volume}{105}}, \bibinfo{pages}{L042202}
  (\bibinfo{year}{2022}).

\bibitem{millan2024triadic}
\bibinfo{author}{Mill{\'a}n, A.~P.}, \bibinfo{author}{Sun, H.},
  \bibinfo{author}{Torres, J.~J.} \& \bibinfo{author}{Bianconi, G.}
\newblock \bibinfo{title}{Triadic percolation induces dynamical topological
  patterns in higher-order networks}.
\newblock \emph{\bibinfo{journal}{PNAS Nexus}} \textbf{\bibinfo{volume}{3}},
  \bibinfo{pages}{270} (\bibinfo{year}{2024}).

\bibitem{zhang2024deeper}
\bibinfo{author}{Zhang, Y.}, \bibinfo{author}{Skardal, P.~S.},
  \bibinfo{author}{Battiston, F.}, \bibinfo{author}{Petri, G.} \&
  \bibinfo{author}{Lucas, M.}
\newblock \bibinfo{title}{Deeper but smaller: Higher-order interactions
  increase linear stability but shrink basins}.
\newblock \emph{\bibinfo{journal}{Science Advances}}
  \textbf{\bibinfo{volume}{10}}, \bibinfo{pages}{eado8049}
  (\bibinfo{year}{2024}).

\bibitem{albert2004structural}
\bibinfo{author}{Albert, R.}, \bibinfo{author}{Albert, I.} \&
  \bibinfo{author}{Nakarado, G.~L.}
\newblock \bibinfo{title}{Structural vulnerability of the north american power
  grid}.
\newblock \emph{\bibinfo{journal}{Physical Review E}}
  \textbf{\bibinfo{volume}{69}}, \bibinfo{pages}{025103}
  (\bibinfo{year}{2004}).

\bibitem{wang2021resilience}
\bibinfo{author}{Wang, S.}, \bibinfo{author}{Gu, X.}, \bibinfo{author}{Luan,
  S.} \& \bibinfo{author}{Zhao, M.}
\newblock \bibinfo{title}{Resilience analysis of interdependent critical
  infrastructure systems considering deep learning and network theory}.
\newblock \emph{\bibinfo{journal}{International Journal of Critical
  Infrastructure Protection}} \textbf{\bibinfo{volume}{35}},
  \bibinfo{pages}{100459} (\bibinfo{year}{2021}).

\bibitem{bian2025unveiling}
\bibinfo{author}{Bian, J.}, \bibinfo{author}{Zhou, T.} \& \bibinfo{author}{Bi,
  Y.}
\newblock \bibinfo{title}{Unveiling the role of higher-order interactions via
  stepwise reduction}.
\newblock \emph{\bibinfo{journal}{Communications Physics}}
  \textbf{\bibinfo{volume}{8}}, \bibinfo{pages}{228} (\bibinfo{year}{2025}).

\bibitem{goldbeter2008biological}
\bibinfo{author}{Goldbeter, A.}
\newblock \bibinfo{title}{Biological rhythms: clocks for all times}.
\newblock \emph{\bibinfo{journal}{Current Biology}}
  \textbf{\bibinfo{volume}{18}}, \bibinfo{pages}{R751--R753}
  (\bibinfo{year}{2008}).

\bibitem{dong2025adaptive}
\bibinfo{author}{Dong, Y.}, \bibinfo{author}{Huo, L.}, \bibinfo{author}{Perc,
  M.} \& \bibinfo{author}{Boccaletti, S.}
\newblock \bibinfo{title}{Adaptive rumor propagation and activity contagion in
  higher-order networks}.
\newblock \emph{\bibinfo{journal}{Communications Physics}}
  \textbf{\bibinfo{volume}{8}}, \bibinfo{pages}{261} (\bibinfo{year}{2025}).

\bibitem{zhao2023occurrence}
\bibinfo{author}{Zhao, D.}, \bibinfo{author}{Li, Y.}, \bibinfo{author}{Liu,
  Q.}, \bibinfo{author}{Zhang, H.} \& \bibinfo{author}{Xu, Y.}
\newblock \bibinfo{title}{The occurrence mechanisms of extreme events in a
  class of nonlinear duffing-type systems under random excitations}.
\newblock \emph{\bibinfo{journal}{Chaos}} \textbf{\bibinfo{volume}{33}}
  (\bibinfo{year}{2023}).

\bibitem{wang2022coherence}
\bibinfo{author}{Wang, Z.}, \bibinfo{author}{Li, Y.}, \bibinfo{author}{Xu, Y.},
  \bibinfo{author}{Kapitaniak, T.} \& \bibinfo{author}{Kurths, J.}
\newblock \bibinfo{title}{Coherence-resonance chimeras in coupled hr neurons
  with alpha-stable l{\'e}vy noise}.
\newblock \emph{\bibinfo{journal}{Journal of Statistical Mechanics}}
  \textbf{\bibinfo{volume}{2022}}, \bibinfo{pages}{053501}
  (\bibinfo{year}{2022}).

\bibitem{zhang2021rate}
\bibinfo{author}{Zhang, X.}, \bibinfo{author}{Xu, Y.}, \bibinfo{author}{Liu,
  Q.}, \bibinfo{author}{Kurths, J.} \& \bibinfo{author}{Grebogi, C.}
\newblock \bibinfo{title}{Rate-dependent bifurcation dodging in a
  thermoacoustic system driven by colored noise}.
\newblock \emph{\bibinfo{journal}{Nonlinear Dynamics}}
  \textbf{\bibinfo{volume}{104}}, \bibinfo{pages}{2733--2743}
  (\bibinfo{year}{2021}).

\bibitem{wang2016levy}
\bibinfo{author}{Wang, Z.}, \bibinfo{author}{Xu, Y.} \& \bibinfo{author}{Yang,
  H.}
\newblock \bibinfo{title}{L{\'e}vy noise induced stochastic resonance in an fhn
  model}.
\newblock \emph{\bibinfo{journal}{Science China Technological Sciences}}
  \textbf{\bibinfo{volume}{59}}, \bibinfo{pages}{371--375}
  (\bibinfo{year}{2016}).

\bibitem{campa2023synchronization}
\bibinfo{author}{Campa, A.} \& \bibinfo{author}{Gupta, S.}
\newblock \bibinfo{title}{Synchronization in a system of kuramoto oscillators
  with distributed gaussian noise}.
\newblock \emph{\bibinfo{journal}{Physical Review E}}
  \textbf{\bibinfo{volume}{108}}, \bibinfo{pages}{064124}
  (\bibinfo{year}{2023}).

\bibitem{holder2017gaussian}
\bibinfo{author}{Holder, A.~B.}, \bibinfo{author}{Zuparic, M.~L.} \&
  \bibinfo{author}{Kalloniatis, A.~C.}
\newblock \bibinfo{title}{Gaussian noise and the two-network frustrated
  kuramoto model}.
\newblock \emph{\bibinfo{journal}{Physica D}} \textbf{\bibinfo{volume}{341}},
  \bibinfo{pages}{10--32} (\bibinfo{year}{2017}).

\bibitem{zhao2023probabilistic}
\bibinfo{author}{Zhao, D.}, \bibinfo{author}{Li, Y.}, \bibinfo{author}{Xu, Y.},
  \bibinfo{author}{Liu, Q.} \& \bibinfo{author}{Kurths, J.}
\newblock \bibinfo{title}{Probabilistic description of extreme oscillations and
  reliability analysis in rolling motion under stochastic excitation}.
\newblock \emph{\bibinfo{journal}{Science China Technological Sciences}}
  \textbf{\bibinfo{volume}{66}}, \bibinfo{pages}{2586--2596}
  (\bibinfo{year}{2023}).

\bibitem{tyulkina2018dynamics}
\bibinfo{author}{Tyulkina, I.~V.}, \bibinfo{author}{Goldobin, D.~S.},
  \bibinfo{author}{Klimenko, L.~S.} \& \bibinfo{author}{Pikovsky, A.}
\newblock \bibinfo{title}{Dynamics of noisy oscillator populations beyond the
  ott-antonsen ansatz}.
\newblock \emph{\bibinfo{journal}{Physical Review Letters}}
  \textbf{\bibinfo{volume}{120}}, \bibinfo{pages}{264101}
  (\bibinfo{year}{2018}).

\bibitem{rajwani2025stochastic}
\bibinfo{author}{Rajwani, P.} \& \bibinfo{author}{Jalan, S.}
\newblock \bibinfo{title}{Stochastic kuramoto oscillators with inertia and
  higher-order interactions}.
\newblock \emph{\bibinfo{journal}{Physical Review E}}
  \textbf{\bibinfo{volume}{111}}, \bibinfo{pages}{L012202}
  (\bibinfo{year}{2025}).

\bibitem{marui2025synchronization}
\bibinfo{author}{Marui, Y.} \& \bibinfo{author}{Kori, H.}
\newblock \bibinfo{title}{Synchronization and its slow decay in noisy
  oscillators with simplicial interactions}.
\newblock \emph{\bibinfo{journal}{Physical Review E}}
  \textbf{\bibinfo{volume}{111}}, \bibinfo{pages}{014223}
  (\bibinfo{year}{2025}).

\bibitem{zhao2025extreme}
\bibinfo{author}{Zhao, D.}, \bibinfo{author}{Li, Y.}, \bibinfo{author}{Liu,
  Q.}, \bibinfo{author}{Kurths, J.} \& \bibinfo{author}{Xu, Y.}
\newblock \bibinfo{title}{Extreme events suppression in a suspended aircraft
  seat system under extreme environment}.
\newblock \emph{\bibinfo{journal}{Communications in Nonlinear Science and
  Numerical Simulation}} \textbf{\bibinfo{volume}{145}},
  \bibinfo{pages}{108707} (\bibinfo{year}{2025}).

\bibitem{liu2023complex}
\bibinfo{author}{Liu, Q.}, \bibinfo{author}{Xu, Y.} \& \bibinfo{author}{Li, Y.}
\newblock \bibinfo{title}{Complex dynamics of a conceptual airfoil structure
  with consideration of extreme flight conditions}.
\newblock \emph{\bibinfo{journal}{Nonlinear Dynamics}}
  \textbf{\bibinfo{volume}{111}}, \bibinfo{pages}{14991--15010}
  (\bibinfo{year}{2023}).

\bibitem{lee2025extreme}
\bibinfo{author}{Lee, S.}, \bibinfo{author}{Kuklinski, L.~J.} \&
  \bibinfo{author}{Timme, M.}
\newblock \bibinfo{title}{Extreme synchronization transitions}.
\newblock \emph{\bibinfo{journal}{Nature Communications}}
  \textbf{\bibinfo{volume}{16}} (\bibinfo{year}{2025}).

\bibitem{suman2024finite}
\bibinfo{author}{Suman, A.} \& \bibinfo{author}{Jalan, S.}
\newblock \bibinfo{title}{Finite-size effect in kuramoto oscillators with
  higher-order interactions}.
\newblock \emph{\bibinfo{journal}{Chaos}} \textbf{\bibinfo{volume}{34}}
  (\bibinfo{year}{2024}).

\bibitem{zhao2022extreme}
\bibinfo{author}{Zhao, D.}, \bibinfo{author}{Li, Y.}, \bibinfo{author}{Xu, Y.},
  \bibinfo{author}{Liu, Q.} \& \bibinfo{author}{Kurths, J.}
\newblock \bibinfo{title}{Extreme events in a class of nonlinear duffing-type
  oscillators with a parametric periodic force}.
\newblock \emph{\bibinfo{journal}{European Physical Journal Plus}}
  \textbf{\bibinfo{volume}{137}}, \bibinfo{pages}{314} (\bibinfo{year}{2022}).

\bibitem{banerjee2020recurrence}
\bibinfo{author}{Banerjee, A.} \emph{et~al.}
\newblock \bibinfo{title}{Recurrence analysis of extreme event like data}.
\newblock \emph{\bibinfo{journal}{Nonlinear Processes in Geophysics
  Discussions}} \textbf{\bibinfo{volume}{2020}}, \bibinfo{pages}{1--25}
  (\bibinfo{year}{2020}).

\bibitem{marwan2023power}
\bibinfo{author}{Marwan, N.} \& \bibinfo{author}{Braun, T.}
\newblock \bibinfo{title}{Power spectral estimate for discrete data}.
\newblock \emph{\bibinfo{journal}{Chaos}} \textbf{\bibinfo{volume}{33}}
  (\bibinfo{year}{2023}).

\bibitem{marwan2023challenges}
\bibinfo{author}{Marwan, N.}
\newblock \bibinfo{title}{Challenges and perspectives in recurrence analyses of
  event time series}.
\newblock \emph{\bibinfo{journal}{Frontiers in Applied Mathematics and
  Statistics}} \textbf{\bibinfo{volume}{9}}, \bibinfo{pages}{1129105}
  (\bibinfo{year}{2023}).

\bibitem{zan2022response}
\bibinfo{author}{Zan, W.}, \bibinfo{author}{Jia, W.} \& \bibinfo{author}{Xu,
  Y.}
\newblock \bibinfo{title}{Response statistics of single-degree-of-freedom
  systems with l{\'e}vy noise by improved path integral method}.
\newblock \emph{\bibinfo{journal}{International Journal of Applied Mechanics}}
  \textbf{\bibinfo{volume}{14}}, \bibinfo{pages}{2250029}
  (\bibinfo{year}{2022}).

\bibitem{menck2013basin}
\bibinfo{author}{Menck, P.~J.}, \bibinfo{author}{Heitzig, J.},
  \bibinfo{author}{Marwan, N.} \& \bibinfo{author}{Kurths, J.}
\newblock \bibinfo{title}{How basin stability complements the linear-stability
  paradigm}.
\newblock \emph{\bibinfo{journal}{Nature Physics}}
  \textbf{\bibinfo{volume}{9}}, \bibinfo{pages}{89--92} (\bibinfo{year}{2013}).

\bibitem{ma2022early}
\bibinfo{author}{Ma, J.}, \bibinfo{author}{Liu, Q.}, \bibinfo{author}{Xu, Y.}
  \& \bibinfo{author}{Kurths, J.}
\newblock \bibinfo{title}{Early warning of noise-induced catastrophic
  high-amplitude oscillations in an airfoil model}.
\newblock \emph{\bibinfo{journal}{Chaos}} \textbf{\bibinfo{volume}{32}}
  (\bibinfo{year}{2022}).

\bibitem{zan2022reliability}
\bibinfo{author}{Zan, W.}, \bibinfo{author}{Jia, W.} \& \bibinfo{author}{Xu,
  Y.}
\newblock \bibinfo{title}{Reliability of dynamical systems with combined
  gaussian and poisson white noise via path integral method}.
\newblock \emph{\bibinfo{journal}{Probabilistic Engineering Mechanics}}
  \textbf{\bibinfo{volume}{68}}, \bibinfo{pages}{103252}
  (\bibinfo{year}{2022}).

\bibitem{lepeu2024critical}
\bibinfo{author}{Lepeu, G.} \emph{et~al.}
\newblock \bibinfo{title}{The critical dynamics of hippocampal seizures}.
\newblock \emph{\bibinfo{journal}{Nature Communications}}
  \textbf{\bibinfo{volume}{15}}, \bibinfo{pages}{6945} (\bibinfo{year}{2024}).

\bibitem{li2025synchronization}
\bibinfo{author}{Li, Z.} \emph{et~al.}
\newblock \bibinfo{title}{Synchronization stability of epileptic brain network
  with higher-order interactions}.
\newblock \emph{\bibinfo{journal}{Chaos}} \textbf{\bibinfo{volume}{35}}
  (\bibinfo{year}{2025}).

\bibitem{baumann2020modeling}
\bibinfo{author}{Baumann, F.}, \bibinfo{author}{Lorenz-Spreen, P.},
  \bibinfo{author}{Sokolov, I.~M.} \& \bibinfo{author}{Starnini, M.}
\newblock \bibinfo{title}{Modeling echo chambers and polarization dynamics in
  social networks}.
\newblock \emph{\bibinfo{journal}{Physical Review Letters}}
  \textbf{\bibinfo{volume}{124}}, \bibinfo{pages}{048301}
  (\bibinfo{year}{2020}).

\end{thebibliography}
\section*{Acknowledgements}
This work is supported by the Key International (Regional) Joint Research Program of the National Natural Science Foundation (NSF) of China under Grant (No. 12120101002). D. Zhao thanks the Sino-German
(CSC-DAAD) Postdoc Scholarship Program.
\section*{Author contributions}
D.Z. and Y.X. conceived and designed the research. D.Z. performed the main calculations and data analysis. J.K. and N.M. contributed to the development of the methodology and the interpretation of the results. D.Z. wrote the original draft of the manuscript. J.K., N.M., and Y.X. supervised the work and critically revised the manuscript. All authors discussed the results and approved the final version of the manuscript.
\section*{Competing interests}
The authors declare no competing interests.
\section*{Additional information}
The online version contains supplementary material. Correspondence and requests for materials should be addressed to D. Zhao or Y. Xu.
\end{document}


\title[Article Title]{Supplementary Information for Synchronization transitions and spike dynamics in a higher-order Kuramoto model with L\'{e}vy noise}


\author[1,2]{\fnm{Dan} \sur{Zhao}}

\author[1,2]{\fnm{J\"urgen} \sur{Kurths}}

\author[1]{\fnm{Norbert} \sur{Marwan}}

\author*[3,4]{\fnm{Yong} \sur{Xu}}\email{hsux3@nwpu.edu.cn}

\maketitle


\section*{Supplementary Results}
\subsection*{Basin of attraction for different coupling parameters}
\begin{figure}
	\centering
	\includegraphics[width=1.1\textwidth]{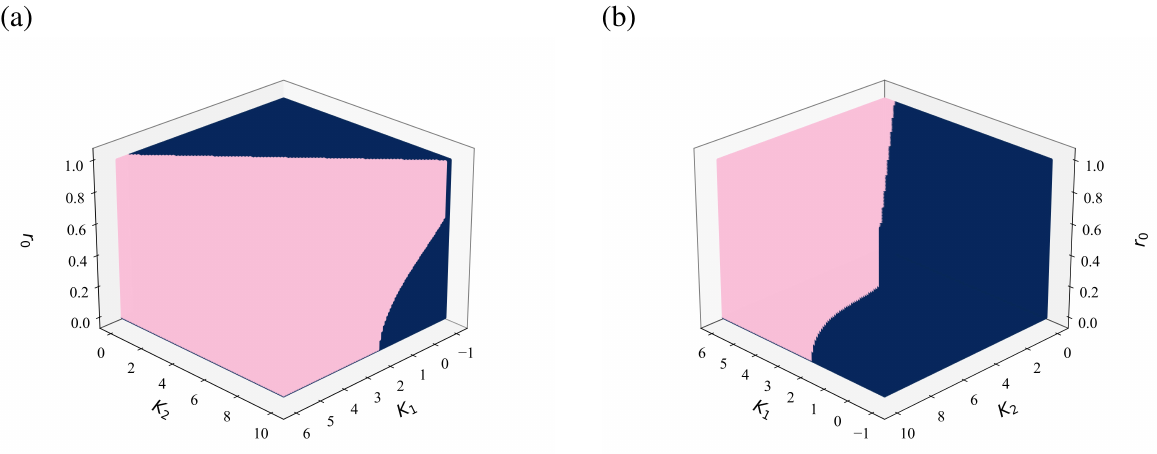}
	\caption{The basin of attraction under 3D-dimensional view. The two subfigures show different angles. The two horizontal axes are $K_1$ and $K_2$, and the vertical axis represents the $r_0$. Synchronization is shown in pink region and incoherence is described in navy region.}
	\label{fig:figSI_2}      
\end{figure}

We consider the basin of attraction for different $K_1$ and $K_2$. Figure \ref{fig:figSI_2} is simply different viewing angles of the same three‐dimensional basin of attraction surface. Figure~\ref{fig:figSI_2} clearly supports the bistable phenomenon from a different perspective, located between incoherence and synchronization. Within this bistable region, initial states with $r_0$ above a critical height on the plane will synchronize, while those below it remain incoherent. 

\subsection*{The response of order parameter and PDF}

\begin{figure}
	\centering
	\includegraphics[width=1\textwidth]{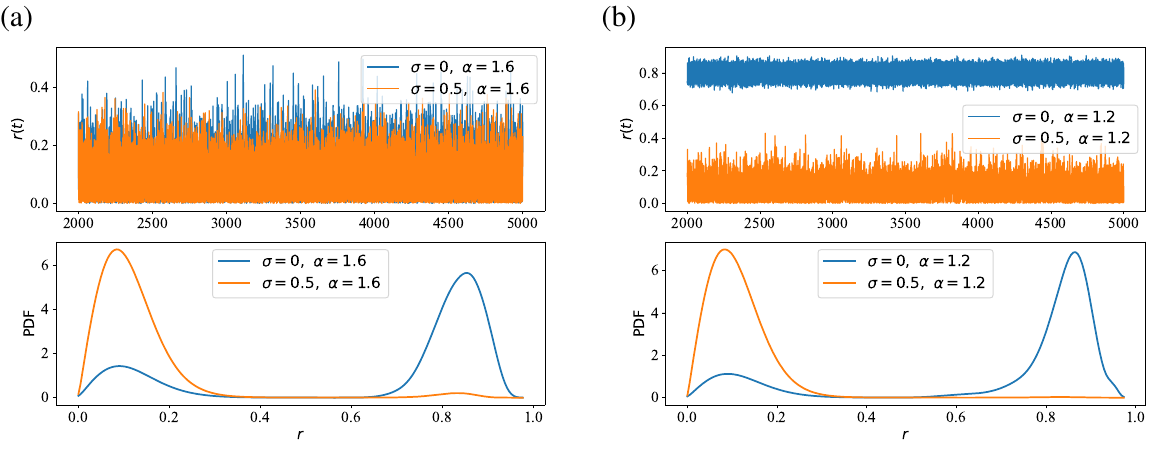}
	\caption{In each subfigure, the first row shows the time series of $r(t)$, and the second row shows the probability density function (PDF) obtained by multiple trajectories. Blue line respresents the deterministic case ($\sigma=0$). Orange line shows the stochastic case. (a) $\sigma=0.5$, $\alpha=1.6$ (orange lines) and $\sigma=0$, $\alpha=1.6$ (blue lines). (b) $\sigma=0.5$, $\alpha=1.2$ (orange lines) and $\sigma=0$, $\alpha=1.2$ (blue lines).}
	\label{fig:figSI_3}      
\end{figure}

To confirm the general influence of $r(t)$ on L\'{e}vy noise, we perform two sets of $\alpha$ and $\sigma$. When $\alpha=1.6$ and $\sigma=0.5$, shown in Fig.~\ref{fig:figSI_3}(a), synchronization state is suppressed directly. The stochastic trajectory (orange curves) dwells near $r=0$, and only rarely climbs toward the synchronized branch, so the PDF peaks sharply at low $r$. In the deterministic case, $r(t)$ never departs significantly from 0 and the PDF plotted below confirms this. It cocurs only oscasionally. As for Fig.~\ref{fig:figSI_3}(b), further increasing tail heaviness virtually eliminates synchronization. The orange curves fluctuate around $r=0$ almost exclusively, and the PDF only shows a small peak near high $r$. Thus, L\'{e}vy noise fundamentally makes the system to transition.

\subsection*{PDF of order parameter}

\begin{figure}
	\centering
	\includegraphics[width=0.6\textwidth]{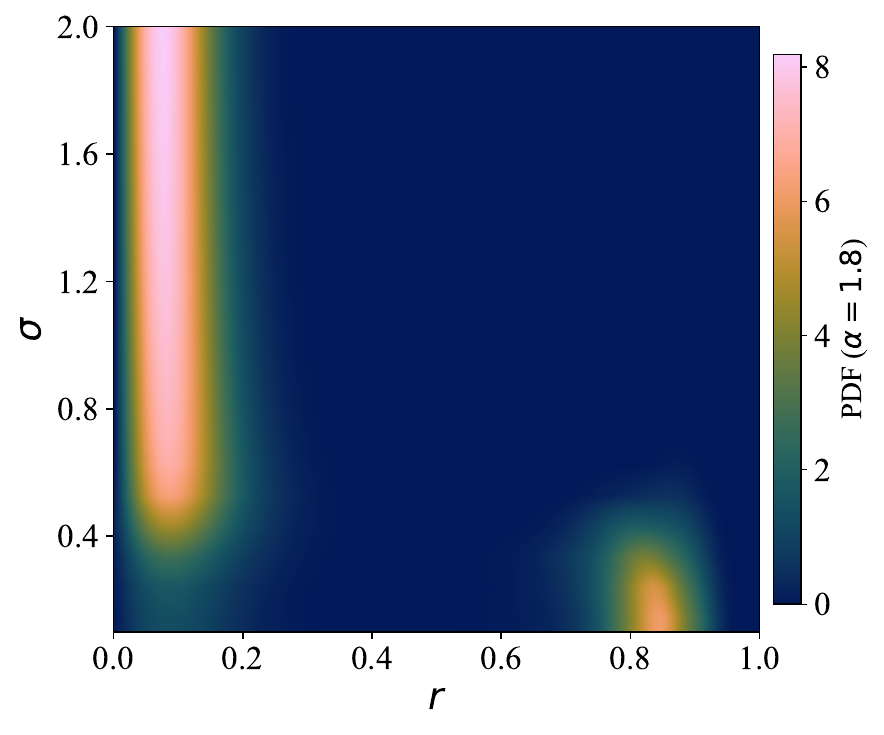}
	\caption{The PDF under different $\alpha$ and $\sigma$. The horizontal axis is the order parameter $r$, the vertical axis is the $\sigma$ and $\alpha$, and different colors represent PDF values. $\alpha=1.8$.}
	\label{fig:figSI_4}      
\end{figure}
To further illustrate how the PDF of the order parameter evolves with $\alpha$, we present the case for $\alpha = 1.8$ in Fig.~\ref{fig:figSI_4}. In comparison with the results shown in Figs.~4(a)(b), the system (1) at $\alpha= 1.8$ remains predominantly in the incoherent state. A synchronization state emerges only within $\sigma<0.4$. This synchronization threshold for $\sigma$ is intermediate between those for the $\alpha=2.0$ (Gaussian) and $\alpha=1.2$ cases, demonstrating that the system's transition behavior evolves continuously as $\alpha$ is varied.

\subsection*{Mean order parameter and mean first-passage time}
To clearly illustrate the effects of $\alpha$ and $\sigma$ on the system (1), we describe the dynamical behaviors of the system (1) by the mean order parameter, where the system (1) may reside in one of two macroscopic states. 
\begin{figure}
	\centering
	\includegraphics[width=1.1\textwidth]{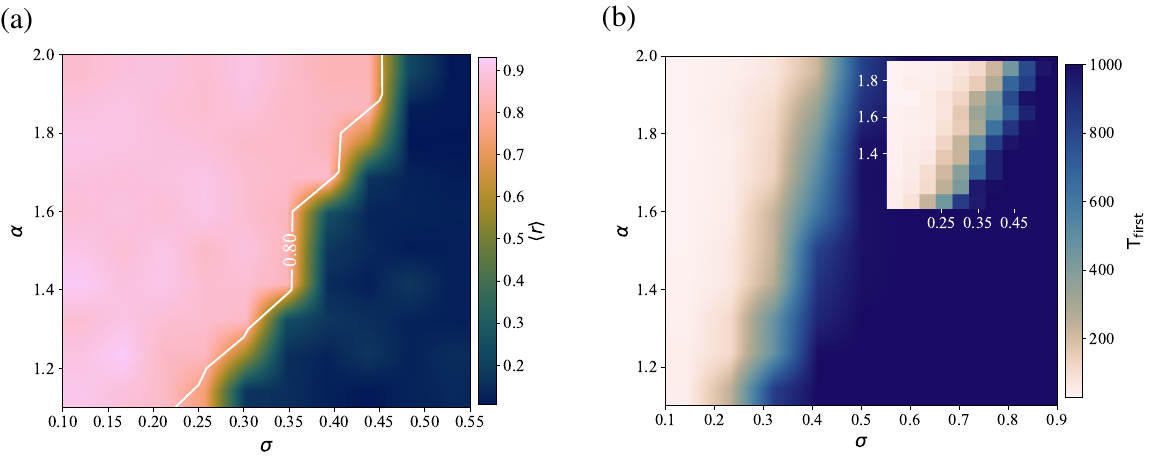}
	\caption{(a) The mean order parameters under different $\alpha$ and $\sigma$. The white curve is the contour line, marking the region with 0.8. (b) Heatmap of the mean first-passage time $\mathrm{T}_{\text{first}}$ under different $\alpha$ and $\sigma$. Dark blue denotes regions where $\mathrm{T}_{\text{first}}$ exceeds our maximum observation window, while brighter regions indicate parameter combinations for which synchronization is reached within finite time. The inset zooms in on $\sigma\in[0.1,0.5]$ and $\alpha\in[1.1,2.0]$.}
	\label{fig:figSI_5}      
\end{figure}
Figure \ref{fig:figSI_5}(a) presents a heatmap of $\left< r \right>$ over $\alpha$ and $\sigma$. The background color indicates the value of $\left< r \right>$, from deep blue ($\left< r \right>=0$, fully incoherent) up to pale pink ($\left< r \right>=1$, fully synchronized), while the superimposed white contour traces the locus $\left< r \right>=0.8$, delineating the boundary between coherent and incoherent regimes. The horizontal axis is $\sigma$ and the vertical axis is $\alpha$.

From the Fig. \ref{fig:figSI_5}(a), we can see that when $\sigma$ is small, $\left< r \right>$ is close to 1, indicating that the system (1) will be synchronized. When $\sigma$ is large, $\left< r \right>$ is close to 0, and the system (1) transitions to an incoherent state. When $\sigma$ is larger than 0.5, $\left< r \right>$ is almost 0, i.e., system (1) is in an incoherent state.
Besides, according to the contour line, the critical values of $\alpha$ and $\sigma$ are very clear. For Gaussian-like fluctuations, such as $\alpha\approx2$, the system (1) tolerates noise up to $\sigma\approx0.5$ before dropping below $\left< r \right>\approx0.8$. Whereas for strongly heavy–tailed noise, such as $\alpha\approx1.2$, the same loss of coherence occurs already at $\sigma\approx0.25$. 

For multi-stable systems, the mean first-passage time (MFPT) is often used to describe how easy it is for a system to transition from one stable state to another, reflecting the stability of the system. 
Figure \ref{fig:figSI_5}(b) provides the $\mathrm{T}_{\text{first}}$ under different $\sigma$ and $\alpha$.  In the heatmap, dark blue denotes that the system (1) remains trapped in the incoherent basin and does not transition to the synchronized state. By contrast, bright areas correspond to $r\geqslant r_{\mathrm{threshold}}$. For small $\alpha$, one needs a smaller $\sigma$ to suppress synchronization. Namely, the boundary between white and blue shifts to the left. As $\alpha$ increases toward 2, the transition between synchronization and incoherence shifts to larger $\sigma$. Besides, increasing $\sigma$ at fixed $\alpha$ always prolongs $\mathrm{T}_{\text{first}}$, eventually saturating in the dark‐blue regime. It becomes easier for the oscillators to overcome synchronization. By contrast, at fixed $\sigma$, increasing $\alpha$ reduces $\mathrm{T}_{\text{first}}$, indicating that Gaussian-like fluctuations are less disruptive to establishing higher-order synchrony than heavy-tailed jumps.

Therefore, weak or Gaussian‐type fluctuations assist the higher‐order coupling in driving the population into synchrony within finite time. By contrast, strong noise effectively prevents the transition to the synchronized attractor.

%
\subsection*{Bifurcation hysteresis}

\begin{figure}
	\centering
		{\includegraphics[width=1.1\textwidth]{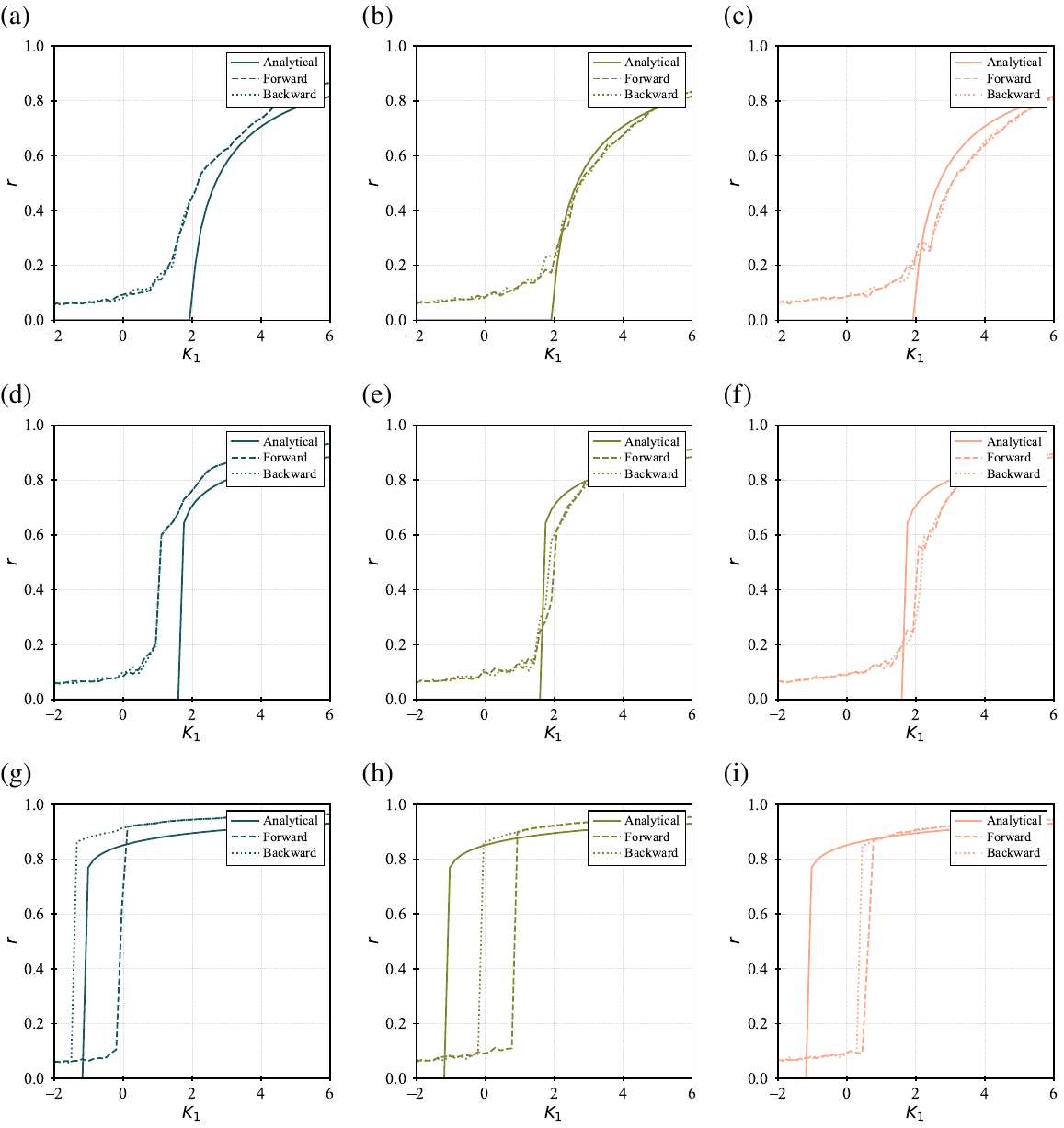}};
	\caption{The bifurcation diagrams under different $K_1$, $K_2$, $\alpha$ and $\sigma$. The solid line is the solution to Eq.~(2). Forward numerical iterations are denoted by long dashed lines with dots, while backward numerical iterations are denoted by short dashed lines. (a)(d)(g) The deterministic case, $\alpha=2$, $\sigma=0$. (b)(e)(h) The Gaussian case, $\alpha=2$, $\sigma=0.5$. (c)(f)(l) The L\'{e}vy noise, $\alpha=1.6$, $\sigma=0.5$. The first row is $K_2=0.0$, the second row is $K_2=4.0$, and the third row is $K_2=10.0$.}
	\label{fig:figSI_9}
\end{figure}

To investigate the effect of the higher-order coupling strength, we investigate the bifurcation diagram for $K_2\in[0.0,4.0,10.0]$, shown in Fig.~\ref{fig:figSI_9}.
For $K_2=0$ in Figs.~\ref{fig:figSI_9}(a)-(c), the forward and backward branches coincide, indicating a single stable incoherent state. At this time, there are only navy scattered points in Fig.~\ref{fig:figSI_2}. As $K_2$ increases to 4.0 and 10.0, shown in Figs.~\ref{fig:figSI_9}(d)-(l), the system (1) exhibits a huge hysteresis. It is sharpest in the Figs.~\ref{fig:figSI_9}(d) and (g), and is almost entirely suppressed in the Figs.~\ref{fig:figSI_9}(f) and (l). Furthermore, both Gaussian and L\'{e}vy noise shift the bifurcation curve to larger $K_1$, with the latter inducing a more pronounced shift.

\subsection*{The definition of spike}

\begin{figure}
	\centering
	\includegraphics[width=0.7\textwidth]{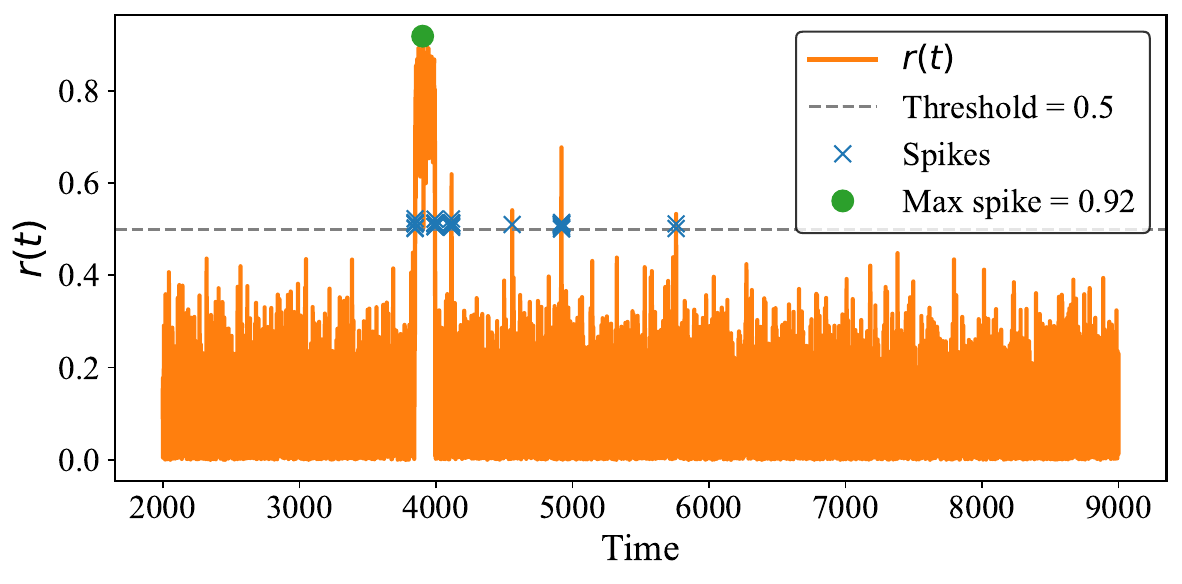}
	\caption{The $r(t)$ under $\alpha=1.4$, $\sigma=0.4$. The dotted line is the threshold $r_{\mathrm{critical}}$, the cross marks the spikes in a given time interval, and the green dot marks the maximum of the spike.}
	\label{fig:figSI_13}      
\end{figure}

To provide an operational definition of spikes, Fig.~\ref{fig:figSI_13} illustrates the responses of the $r(t)$. A spike is defined as any event where the $r(t)$ surpasses a pre-defined threshold, indicated by the dashed line. The blue crosses in the Fig.~\ref{fig:figSI_13} mark some spikes. The green dot highlights the spike with the maximum amplitude. 

